\documentclass[preprint, 12pt,3p]{elsarticle}

\usepackage{amssymb}

\usepackage{amsmath}
\usepackage{graphicx}
\usepackage{epstopdf}
\usepackage[tight,footnotesize]{subfigure}
\usepackage{units}
\usepackage{flushend}
\usepackage[justification=centering]{caption}
\usepackage{color, colortbl}
\usepackage[table]{xcolor}
\usepackage{multirow}
\usepackage{times}
\usepackage{rotating}
\usepackage{booktabs}
\usepackage{indentfirst}
\usepackage{bm}
\usepackage{xfrac}
\usepackage[ruled,vlined,linesnumbered]{algorithm2e}

\usepackage{colortbl} 
\usepackage{hhline} 
\usepackage{multirow}

\journal{Computer Communications}

\begin{document}

\begin{frontmatter}

\title{Towards a Distributed and Infrastructure-less Vehicular Traffic Management System}

\author[label1]{Ademar T. Akabane\corref{cor1}}
\ead{takeo@lrc.ic.unicamp.br}
\ead[url]{http://www.lrc.ic.unicamp.br/~takeo/}
\author[label1]{Roger Immich}
\ead{roger@ic.unicamp.br}
\author[label1]{Luiz F. Bittencourt}
\ead{bit@ic.unicamp.br}
\author[label1]{Edmundo R. M. Madeira}
\ead{edmundo@ic.unicamp.br}
\author[label1]{Leandro A. Villas}
\ead{leandro@ic.unicamp.br}
\address[label1]{Institute of Computing, University of Campinas, Campinas-SP, Brazil}

\cortext[cor1]{I am corresponding author}

\begin{abstract}
In the past few years, several systems have been proposed to deal with issues related to the vehicular traffic management. Usually, their solutions include the integration of computational technologies such as vehicular networks, central servers, and roadside units. Most systems use a hybrid approach, which means they still need a central entity (central server or roadside unit) and Internet connection to find out an en-route event as well as alternative routes for vehicles. It is easy to understand the need for a central entity because selecting the most appropriate vehicle to perform aforementioned procedures is a difficult task. This is especially true in a highly dynamic network. In addition to that, as far as we know, there are very few systems that apply the altruistic approach (not selfish behavior) to routing decisions. Because of that, the issue addressed in this work is how to perform the vehicular traffic management, when an en-route event is detected, in a distributed, scalable, and cost-effective fashion. To deal with these issues, we proposed a distributed vehicle traffic management system, named as dEASY~(\textbf{d}istributed v\textbf{E}hicle tr\textbf{A}ffic management \textbf{SY}stem). The dEASY system was designed and implemented on a three-layer architecture, namely \textit{environment sensing and vehicle ranking}, \textit{knowledge generation and distribution}, and \textit{knowledge consumption}. Each layer of the dEASY architecture is responsible for dealing with the main issues that were not addressed in related works or could be improved. The essential task of each the layer is: the first layer deals with the task of selecting the most appropriate vehicle, the second one addresses the knowledge generation and distribution, and in the last layer is applied an altruistic approach to choose an alternative route. Simulation results have shown that, compared with other systems from the literature, our proposed system has lower network overhead due to applied vehicle selection and broadcast suppression mechanisms. In average, dEASY also outperformed all other competitors in what regards to the travel time and time lost metrics. Through the analysis of results, it is possible to conclude that our infrastructure-less system is scalable and cost-effective. 
\end{abstract}

\begin{keyword}
Vehicle Traffic Management System \sep Altruistic Rerouting \sep Vehicular ad hoc Networks \sep Infrastructure-less \sep Environment Sensing \sep Knowledge Generation and Distribution \sep Knowledge Consumption
\end{keyword}

\end{frontmatter}


\section{Introduction}

Traffic congestion is a daily occurrence for citizens living in large cities around the world. This problem tends to worsen with the economic and population growth in the urban centers. The increasing vehicular traffic demand may overwhelm the existing transport infrastructure, especially during rush hours~\cite{choudhary2016urban,akabane2017apolo}. To improve on this issue, two immediate solutions come to mind: \textit{(i)} the expansion of road infrastructure; or \textit{(ii)} the amendment of the traffic management system. In the former solution, the cost of road infrastructure expansion is often impractical, due to financial and/or physical-space constraints. The latter solution, on the other hand, allows the use of already existing technologies, along with the new ones, to improve the efficiency of the traffic management system. This can be done without the need to invest in new road infrastructure. Due to the limitations mentioned in the first solution, significant research efforts have been directed toward the second one~\cite{nellore2016survey,doolan2017ecotrec, wang2016next, wang2015real, pan2017divert}. One of the promising technologies for an efficient traffic management system is the Vehicular ad hoc Networks~(VANETs).

The highly dynamic topology of VANETs is one of the features that sets it apart from the other types of networks. Besides that, the mobility pattern of the mobile nodes (vehicles) in these networks is restricted by the road’s pathways as well as by the traffic conditions~\cite{akabane2017apolo}. In VANETs, all external interaction is done through wireless communication links either between vehicles~(vehicle-to-vehicle, V2V) or between the vehicle and the roadside unit~(vehicle-to-infrastructure, V2I)~\cite{akabane2015, akabane2017applying}. The communication technology commonly employed follows the WAVE (Wireless Access for Vehicular Environment - 802.11p) standard. WAVE supports seven channels at the 5.9 GHz band~\cite{dsrc2016, etsi2014302}, where one of them is exclusively dedicated to the control channel, and the other six are dedicated to service channels~\cite{dsrc2016}.

Most VANETs applications require to be aware of the local situation to find out en-route events (e.g. traffic congestion)~\cite{hartenstein2008, hartenstein2009}. One way to get such awareness is to apply the beaconing approach~\cite{schmidt2010}. This approach performs the periodic exchange of one-hop beacon messages through the control channel. As a result, each vehicle will be aware of information from neighboring vehicles that are within its transmission range. The content of the beacon message is associated with the GPS information such as location coordinates, speed, direction, as well as other vehicle information~\cite{schmidt2010}. 

Using the beaconing approach, a real-time traffic management system can collect raw data and extract useful information~(or knowledge) about the traffic congestion in a region of interest. Several systems were designed and implemented to recommend an alternative route, in real-time, when an en-route event is detected~\cite{doolan2017ecotrec, wang2016next, wang2015real, pan2017divert}. 

All the systems mentioned above apply a hybrid approach. In other words, the raw data collection task and knowledge extraction are done by a central entity~(central server or roadside units), and the calculation of the alternative route is performed by the vehicles. It is easy to understand the use of the infrastructure for such tasks since selecting the best-located vehicle in VANETs is not a simple task due to its highly dynamic topology. Furthermore, the previously mentioned systems do not provide a procedure to deal with the broadcast storm problem as well, which takes place during the knowledge dissemination process. Neglecting this problem, the scalability of systems may be compromised. This is especially true in regions where there is a high concentration of vehicles. Another important aspect, when it comes to alternative route planning in the traffic management system, is the altruistic approach. It helps to select a better route by taking into account the neighbor's vehicles routes during the planning phase. In all abovementioned systems, only one applied the altruistic approach~\cite{pan2017divert}. Although several systems have been proposed for vehicular traffic management, there is still a long way to go to implement these systems in practice.

Motivated by the aforementioned concerns, This work proposes an infrastructure-less system for traffic management based on three-layer architecture named dEASY~(\textbf{d}istributed v\textbf{E}hicle tr\textbf{A}ffic management \textbf{SY}stem). The three layers are \textit{environment sensing and vehicle ranking}, \textit{knowledge generation and distribution}, and \textit{knowledge consumption}. In a bottom-up fashion, the first layer, \textit{environment sensing and vehicle ranking}, applies a novel vehicle ranking mechanism that was inspired in Google's PageRank~\cite{page1999pagerank}. The proposal of this innovative mechanism allows the vehicle to autonomously compute its score based on the number of one-hop communication links. The main objective here is to use the ranking score to select the most appropriate vehicle for tasks to be performed on the next layer. In addition to that, we use the beaconing approach to be aware of the local situation. In the second layer, \textit{knowledge generation and distribution}, the vehicle with the highest ranking score processes the raw data received to extract knowledge about the condition of the traffic congestion. Once the knowledge generation is completed, it will be disseminated. During this process, a broadcast suppression mechanism is applied to avoid any unnecessary network overhead. In the last layer, \textit{knowledge consumption}, vehicles use the received knowledge, in addition to the neighborhood route information, to create an altruistic alternative route that avoids the congestion spot without creating others.

In order to validate our proposal, the simulation-based evaluation has been analyzed from three perspectives: \textit{(i)} control channel assessment; \textit{(ii)} scalability assessment; and \textit{(iii)} traffic management assessment. The dEASY system outperformed all its competitors in all the analyzed perspectives. Because of that, it is possible to assert that dEASY is a scalable and cost-effective system for real-time vehicle traffic management.

The main contributions of this paper can be summarized as follows:

\begin{itemize}
  \item The proposal of a scalable, distributed three-layer system for real-time traffic management (Section \ref{vehicle_traffic_management_system});
  \item A novel vehicle ranking mechanism (Section \ref{egocentric}). The proposed mechanism allows selecting the most relevant vehicle in the VANETs, as demonstrated in the scalability assessment (Section \ref{scalability_assessment}); 

  \item An altruistic rerouting approach along with the entropy-based shortest path (Section \ref{consumption_knowledge_layer}). This approach enables the computation of an alternative route and has demonstrated as an interesting option in the traffic management assessment (Section \ref{traffic_management_assessment}).
\end{itemize}

The remainder of this paper is organized as follows. Section II provides a brief review of the related work. Section III details our distributed three-layer system for real-time traffic management. Simulation setup is described in Section IV. Numerical results and analysis are presented in Section V. Section VI concludes this work and looks out on future improvements.

\section{Related Work}

Many works have proposed congestion detection systems and real-time vehicle path planning with the support of VANETs. They also use a combination of different technologies such as induction loops, central server, and road sensors to achieve their goals~\cite{doolan2017ecotrec, wang2016next, wang2015real, pan2017divert, liu2017high, kumar2018ant, wang2018optimizing}. The main purpose of these systems is redistributing the flow of vehicle traffic that is going to the congested area, using real-time traffic data collected, to the non-congested area. 

Several emerging technologies provide a large opportunity to decrease vehicular traffic congestion, and also the emission of polluting gases, by only monitoring traffic conditions. Based on this, Liu et al.~\cite{liu2017high} proposed a four-tier centralized architecture for urban traffic management combined with 5G wireless network, SDN (software-defined networks), and MEC (mobile edge computing) technologies. The proposed tiers are the environment sensing layer, the communication layer, the MEC server layer, and the remote core cloud server~(RCCS) layer. The environment sensing layer is responsible for the perception of real-time traffic conditions. Vehicles actively report their status as well as environmental data to RSUs. The communication layer is composed of two emerging network paradigms, namely 5G and SDN. While the 5G network is responsible for direct communication between vehicles and infrastructures, SDN is responsible for decoupling network control and forwarding functions, enabling thus network control to become programmable. The MEC server layer is introduced to improve the responsiveness for the traffic congestion system. It is deployed on the roadside infrastructure in order to keep it close to the end user. The remote core cloud server layer processes all of the traffic data to identify the congestion points. This layer is also responsible for informing vehicles and RSUs about the events monitored by the system in the four tiers.

Another example of an architecture for traffic management is the Ecotec~\cite{doolan2017ecotrec}. Its main objective is to reduce gas emissions along the vehicle's route without significantly increasing their travel times. EcoTrec relies on the periodic exchange of information about road characteristics and traffic conditions in the route segment. Based on the information collected, it applies a fuel consumption model that takes into account the vehicle route to build and recommend alternative routes. The vehicles periodically send information about the monitored environment to the nearest RSU. With this information, the utility function computes and updates the optimum route. The function's goal is the best avail of each road potential. The vehicles' routes are planned according to the shortest path algorithm with basic load balancing strategy. This strategy is applied to avoid the most popular roads to become congested. EcoTech architecture is constituted of three components: \textit{Vehicle Model}, \textit{Road Model}, and \textit{Traffic Model}. The \textit{Vehicle Model} is built and updated by each vehicle individually, using information from the GPS sensors and accelerometer. On the other hand, the vehicle's local traffic conditions are applied to build and maintain the \textit{Traffic Model}. Both the \textit{Traffic Model} and the \textit{Road Model} are maintained at a central server and updated with information on the nearby roads around the RSU. Both works~\cite{doolan2017ecotrec,liu2017high} require RSUs, but it is not clear how they are distributed in the scenario. In addition to that, a broadcast suppression mechanism was not applied during the information dissemination process. This can affect the system performance and scalability. Another gap is the absence of the scalability assessment of the system.

Another example of a centralized vehicle rerouting system is the Next Road Rerouting (NRR)~ \cite{wang2016next}. The NRR's goal is to assist drivers in making the most suitable next road choice by focusing on a higher travel time reliability in the face of congestions. The system is deployed as an add-on to the typical 3-tier architecture of SCATS (Sydney Coordinated Adaptive Traffic System)~\cite{sims1980sydney}. On top of this architecture is the Traffic Operation Center. It can manage up to 64 regional computers residing in the middle tier. Each computer is responsible for coordinating the synchronization of all traffic lights’ phases in its region, based on the real-time traffic information gathered from loop detectors. At the bottom tier are the intersections (up to 250), where the traffic lights and loop detectors are deployed. The rerouting calculation is done by the regional computer that will recommend only the next road with the least cost for each rerouting request. All requests are done through V2I communication. The routing cost applied in NRR is expressed in terms of road occupancy, travel time estimation, geographic distance to destination, and geographic closeness of congestion. After the vehicle enters the suggested optimal next road, it calculates the remaining route with the aid of the vehicle navigation system. In this work, it is not clear how the sensing of the environment is performed to detect the en-route event. In addition to that, no experimental evaluation regarding the scalability of the system was presented. 

A real-time global path-planning system has been proposed by Wang et al.~\cite{wang2015real}. This system uses both VANETs and public transportation system to enable real-time communications among vehicles, RSUs, and a vehicle-traffic server. In this case, taxis and buses are considered as super-nodes, and they can directly upload the received warning message to the nearest cellular base station (BS). The BS, in turn, will deliver the message to the vehicle-traffic server. Moreover, it is assumed that there is an RSU placed at each intersection. All the vehicles periodically exchange information about their movement. When the congestion is detected, the vehicles around the RSU will generate and forward the warning message to other vehicles through V2V communication. To reduce the redundancy of multi-hop relaying, only taxis/buses that are within the transmission range will continue the data dissemination process. Otherwise, the farthest vehicle in the same lane will be chosen as the next relay. Every time an RSU or a cellular BS receives a warning message, it forwards the message to the traffic server through the wired network. After receiving it, the server will perform the path-planning algorithm based on the road traffic information. Last, after the server concluded the path planning, the replanned paths are forwarded back to vehicles by the reverse way, i.e., traffic server, RSU, and the vehicle in need of alternative path. The path-planning algorithm takes into account both the traffic flows of the network and the path-planning cost. The main goal of this algorithm is to maximize the spatial utility while minimizing travel cost. This study, as in work by Doolan et al. \cite{doolan2017ecotrec}, does not evaluate the scalability of the system. This evaluation is important because there is a high number of messages being exchanged in the network. In addition to that, it is not clear how the identification of congestion spots is performed.

DIVERT is a hybrid vehicular rerouting system for congestion avoidance~\cite{pan2017divert}. This proposal is considered hybrid because it uses a central server, reachable over the Internet communication, to provide a global view of the traffic situations. The central server operates as a coordinator that collects location reports, identifies traffic congestion spots, and distributes rerouting notifications to the vehicles. Each vehicle estimates local density using a periodic exchange of beacon packets among its neighbors, and sends this estimate to the server that handles the task of identifying the congestion. When points of congestion are identified, the central server sends the traffic map to the vehicles that have sent the latest updates. Afterward, these vehicles distribute the traffic map received from the central server in their region of interest. Such system offloads the selection process of the alternative route to the vehicles, thus the rerouting process becomes practical in real-time. The selection of an alternative route, for each vehicle affected by congestion, is given by the $k$ loop-less shortest paths algorithm based on the current position and destination. In addition to that, DIVERT takes collaborative rerouting decisions, i.e., the vehicles situated in the same region exchange information about their alternate routes over V2V communications. Through collaborative rerouting, vehicles can create an awareness of traffic conditions which could contribute to the choice of the alternative route, thus avoiding additional congestion. This work lacks a broadcast suppression mechanism during the two distinct steps of the information distribution (traffic map and alternative route), which can affect the scalability of the system. In addition, the difficulty in understanding the system scalability may in part be attributed to the lack of a detailed performance evaluation of the DIVERT's system.

The work in \cite{kumar2018ant} proposes a vehicular traffic control system that applies the IoV~(Internet of Vehicles) to prevent heavy traffic formation and accidents. To this end, the proposed traffic control system segments the evaluated scenario into small sub-scenarios. In each sub-scenario, there is an infrastructure available that collects and processes all the traffic data. If a congestion event occurs, it applies the ant colony algorithm in order to find the optimal route for vehicles that are going towards the congestion spot. Likewise, in \cite{wang2018optimizing} CDRAM~(Content Dissemination framework for Real-time trAffic Management) is proposed applying IoV. CDRAM is composed of three main components: RSU, BS, and TMS~(traffic management system). The RSU component is used to upload the traffic data generated by vehicles to TMS. The BS component can provide full coverage of wireless communications for users in urban areas, and it is used when RSUs are unavailable. Finally, TMS receives and processes all traffic data in order to identify the traffic congestion spots. In addition, TMS computes and informs alternative routes for vehicles. This allows them to avoid congested areas. Similarly to the works by Doolan et al.~\cite{doolan2017ecotrec} and Wang et al.~\cite{wang2015real}, the scalability of the system has not been evaluated. This raises a number of concerns with respect to its real applicability in a cost-effective way.

Due to the highly dynamic network conditions in VANETs, some systems choose to employ an infrastructural approach, thus eliminating the difficult task of selecting the most relevant vehicles to identify congestion and recommend alternative routes.
Such type of infrastructure, with central server or RSU, performs data aggregation in order to extract knowledge about traffic congestion spots. 
In addition to that, some solutions implicitly assume that there are RSUs available along with all the extension of public roads. This approach may be impractical, particularly when a specific region has no pre-existing infrastructure. Based on the challenges identified in this work, we propose the dEASY system, which is an infrastructure-less system for vehicle traffic management based on a three-layer architecture.

\section{Distributed and Infrastructure-less Vehicular Traffic Management System} \label{vehicle_traffic_management_system}

In this section, we first present the assumptions used in our work and then detail our infrastructure-less system for vehicle traffic management.

\subsection{Assumptions}

A number of assumptions are made in order to build our vehicle traffic management system. The key assumptions are as follows:

\begin{enumerate}
  \item Each vehicle has a GPS receiver to provide the primary navigation data;
  \item Each vehicle has a bidirectional communication link with neighbor vehicles within the transmission range; 
\item The communication link breaks if the distance between vehicles is greater than the transmission range;
  \item All vehicles have the same transmission range;
  \item The propagation model employed is the Two-Ray Path Loss.
\end{enumerate}

\subsection{System Overview}

\begin{figure*}[h!]
\centering
  \captionsetup{justification=centering}
  \includegraphics[width=1\textwidth]{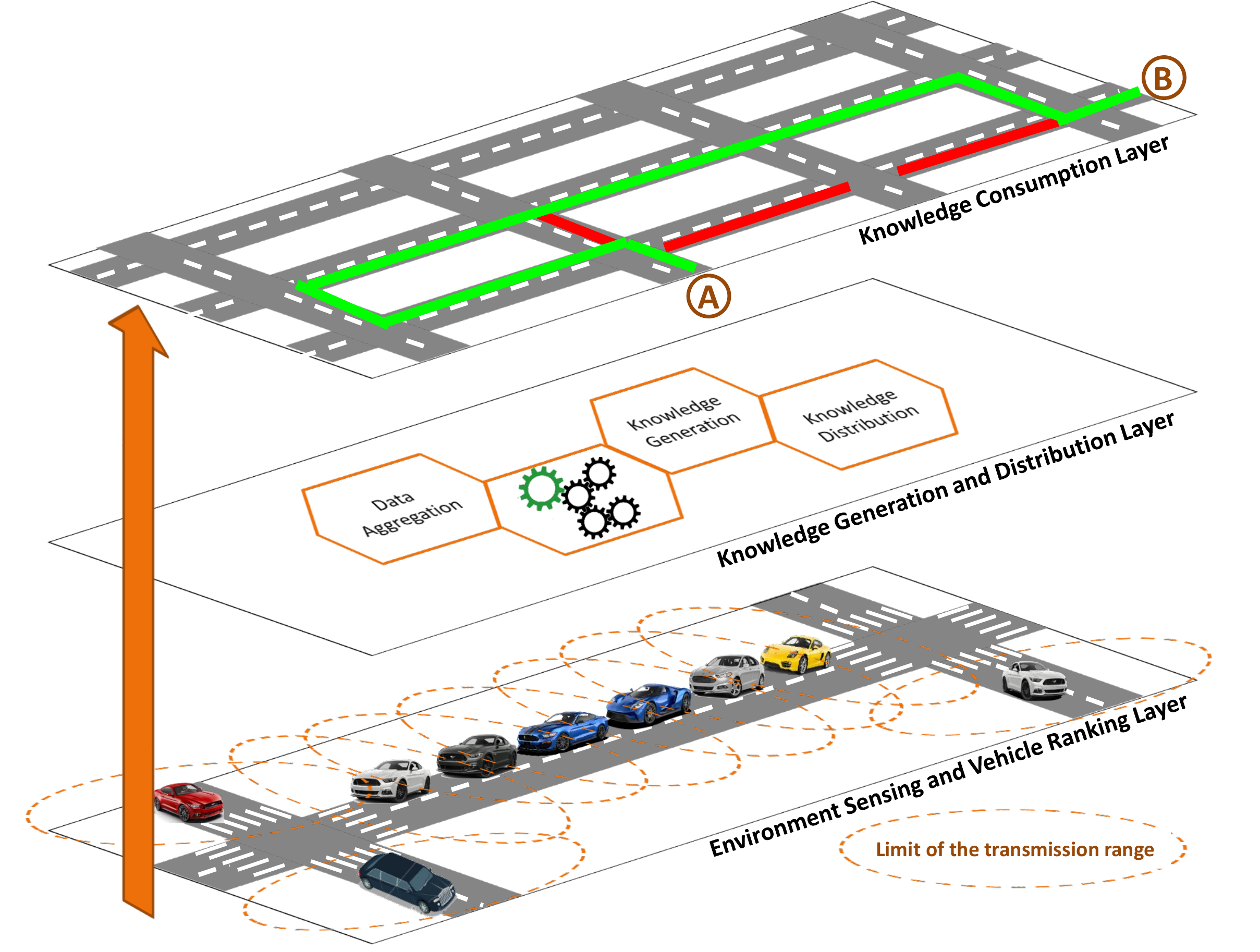}
  \caption{The dEASY's three-layer system architecture.}
\label{architecture}
\end{figure*}

Figure~\ref{architecture} depicts the dEASY's three-layer system architecture for traffic management. In a bottom-up view, the bottommost layer depicts the environment sensing and vehicle rank. The mid layer depicts knowledge generation and distribution, and the upmost layer depicts knowledge consumption. In the following, we will introduce an overview of how literature systems have addressed in each of these three layers. After that, in each subsection, we show in details each layer of the dEASY's three-layer system architecture proposed in this work.

The \textbf{Environment Sensing Layer} plays a significant role in raw data collection, in addition to that is possible to extract from it an accurate knowledge about traffic conditions of the roads. This layer relies on V2V and V2I communications as well as roadside sensors and transportation infrastructure to acquire real-time raw data traffic to build a knowledge base and keep it up to date. To this end, vehicles share a beacon message periodically between them or with the central entity. This type of message typically contains data gathered directly from the global navigation satellite system (GNSS) receiver, which includes the current position, speed, and heading, just to name a few. 

The \textbf{Knowledge Generation and Distribution Layer} includes the knowledge-generation processes such as processing and aggregation of the raw data. This step is vital to meet the different needs according to the context of the service. In this process, the individual raw data, provided by the bottommost layer, is gathered and grouped until new knowledge emerges from this mass of data. Additionally, this layer is also responsible for delivering services to customers such as route suggestion and en-route events.

The \textbf{knowledge Consumption Layer} is responsible for providing a knowledge-based decision making procedure. This decision can be approached in two different ways: \textit{(i)} the decisions are made to benefit the overall system - \textit{altruistic approach}; or \textit{(ii)} the decisions are made only seeking self-benefit - \textit{selfish approach}. For a better understanding of the abovementioned approaches, let us take as an example a congestion avoidance service. As soon as any congestion is detected, vehicles traveling towards it should search for an alternative route. However, by applying a selfish approach, they can create a secondary congestion, particularly when vehicles in the congested roads have similar destinations. In this situation, it is important that vehicles share their decision-making process with the maximum number of participants involved in the congestion. This allows a better global decision-making process for all the participants, in addition to a better overall network flow. Each one of these layers is described in detail below.

\subsection{Environment Sensing and Vehicle Ranking} \label{egocentric}

The environment sensing involves collecting data about the driving surroundings. Each vehicle periodically broadcasts its current GNSS data through beacon messages to all the vehicles within the transmission range. After receiving the beacon messages, vehicles can perform data aggregation and broadcast the result in the network. However, if this is done by all participating vehicles in an uncoordinated fashion, it will consume the entire network bandwidth in a short time. Based on that, it is useful to identify the best-located vehicles in the network continually since the network is highly dynamic and the localization of vehicles is time-dependent. 

Identifying the best-located vehicles in a VANET is a very challenging task due to its highly dynamic topology. On the other hand, once they are identified, it can beneficial for a large number of services, such as the ones that spread the information flow through the network~\cite{daly2007}.

It is known that Google's PageRank~\cite{page1999pagerank} algorithm ranks the importance of websites based on the number of web-links directed towards it. Therefore, the higher the number of links, the greater the popularity of the page on Internet searches. Inspired by this idea, we propose an innovative vehicle ranking mechanism, named $V_{rank}$. This mechanism enables the vehicles to autonomously compute their score based on the number of communication links established with their neighbors. Another $V_{rank}$ feature is the joint use of the radio propagation model, as can be observed in Equation~\ref{VRank}.

\begin{equation}
  \label{VRank}
  V_{rank} = \alpha EBM + (1-\alpha) RPM
\end{equation}
where $EBM$ and $RPM$ are the Egocentric Betweenness Metric and the Radio Propagation Model, respectively. The aim is to improve the process of data propagation, among vehicles, through a path with minimum interference. Both parameters will be explained in more details later. The weighting factor, $\alpha$, is an indicator of the importance of each parameter for the calculation of vehicle relevance in the network, where $\alpha$ $\in$ (0,1). 

The EBM applies a centrality-based social-popularity approach inspired in social networks analysis (SNA) technique. It analyzes the social data structure and social relation to understand the characteristics of a network~\cite{Yang2008social, scott2017social}. There is a key difference between the traditional SNA and the EBM. The first one usually requires the entire knowledge of the network topology to perform the analysis~\cite{Yang2008social, marsden2002}, while in the second one the analysis is done over the structure of ego-networks, and requires only the local knowledge~\cite{marsden2002, everett2005}. Therefore, applying the structure of ego-networks to study its behavior is an attractive approach in VANETs. Next, we introduce the formal definition of the EBM and how it is calculated.

\begin{figure}[h!]
\centering
  \captionsetup{justification=centering}
  \includegraphics[width=0.25\textwidth]{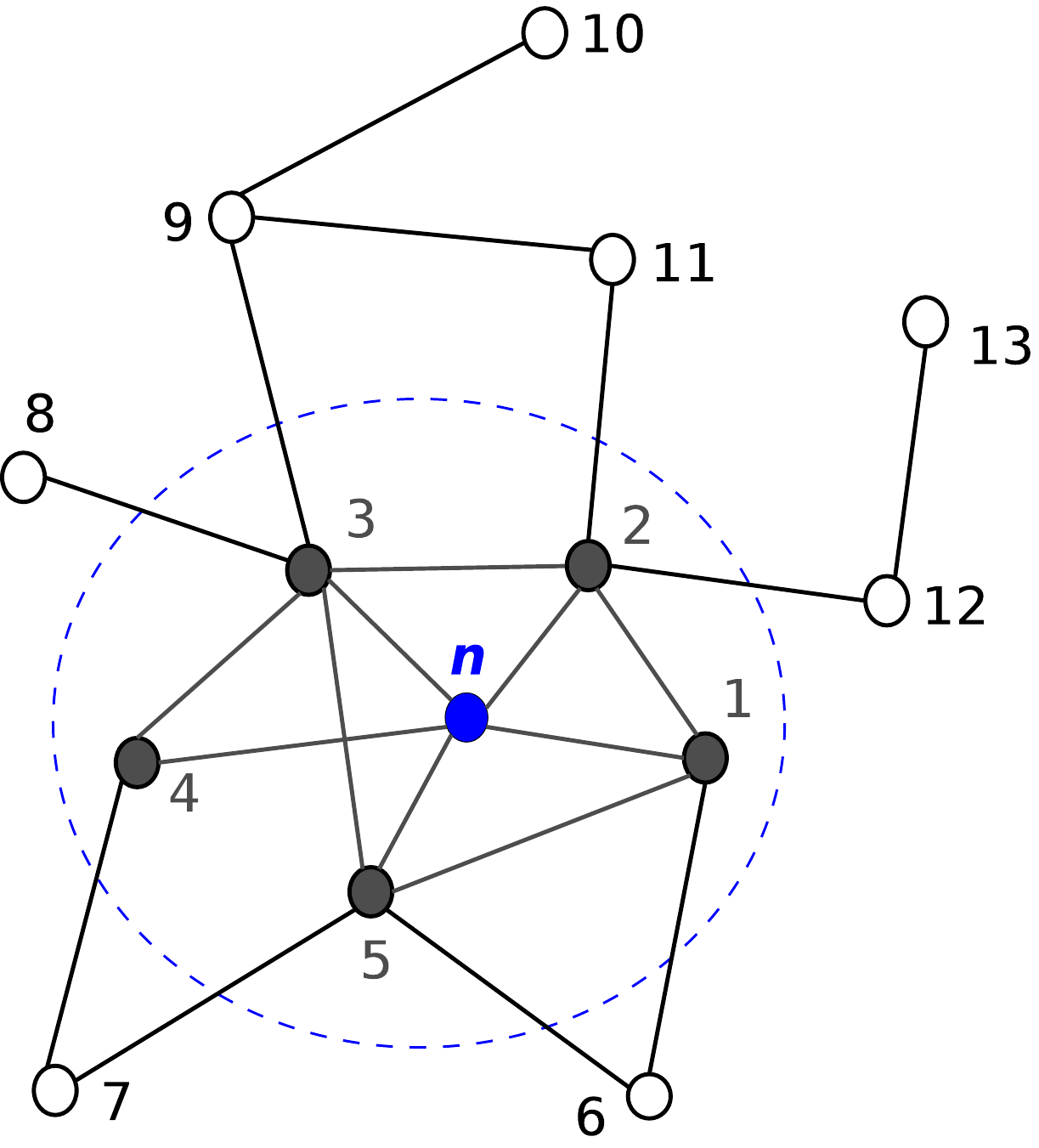}
  \caption{An illustrative example of the ego-network.}
\label{egoNetworks}
\end{figure}

\vspace{0.1cm}
\noindent
\textit{\textbf{EBM} is computed using an ego-network representation. By definition, an ego-network is a subnetwork constituted of a single node (ego - represented by label $n$ in Figure~\ref{egoNetworks}) together with the nodes to which they are connected to (alters - nodes 1, 2, 3, 4, and 5), and all links between alters~\cite{marsden2002,leskovec2012}. Let $N^{r}_{n}$ be the set of nodes, $v'$, that are r-hop away from $n$ (ego), i.e., $N^{r}_{n} = \{v'\in V \wedge 1 \leq d(n,v') \leq r\}$, where $d(n,v')$ denotes a one-hop link between $n$ and $v'$. Thereby, $1^{th}$-order of node $n$ consists of undirected graph $G^{1}_{n} = (V^{1}_{n}, E^{1}_{n})$, where the set of nodes corresponds to $V^{1}_{n} = \{N^{1}_{n} \cup \{n\}\}$ and the set of edges corresponds to $E^{1}_{n} = \{(i,j) \in E^{1}_{n}|i,j \in V^{1}_{n}\}$.}

Mathematically, node-to-node links can be represented by a symmetric adjacency matrix A ($k \times k$), where $k$ is the number of one-hop neighbor nodes. Thereby, each element in A, $a_{i,j}$, can be given by: 

\vspace{-0.4cm}
\[ 
  a_{ij}=
  \begin{cases}
    1	& \quad \text{if there is a direct link between $i$ and $j$}\\
    0	& \quad \text{otherwise}\\
  \end{cases}
\]

Therefore, the EBM of a certain node, $n$, can be calculated by the sum of the reciprocal values of the $A_n^2[1-A_n]_{i,j}$, as defined in the Equation~\ref{EBM}~\cite{everett2005}.

\begin{equation}
  \label{EBM} 
  EBM_{(n)}=\sum_{\substack{A_n(i,j) \neq 0, i < j}}\frac{1}{A_n^2[1-A_n]_{i,j}}
\end{equation}
where $A_n$ depicts the adjacency matrix of the node $n$, 1 is a matrix of all 1's, and the matrix $A_n^2$ provides the number of geodesic distances of length 2 between the node pairs $i$ and $j$. 

\begin{figure}[htp!]
\centering
  \captionsetup{justification=centering}
  \includegraphics[width=0.4\textwidth]{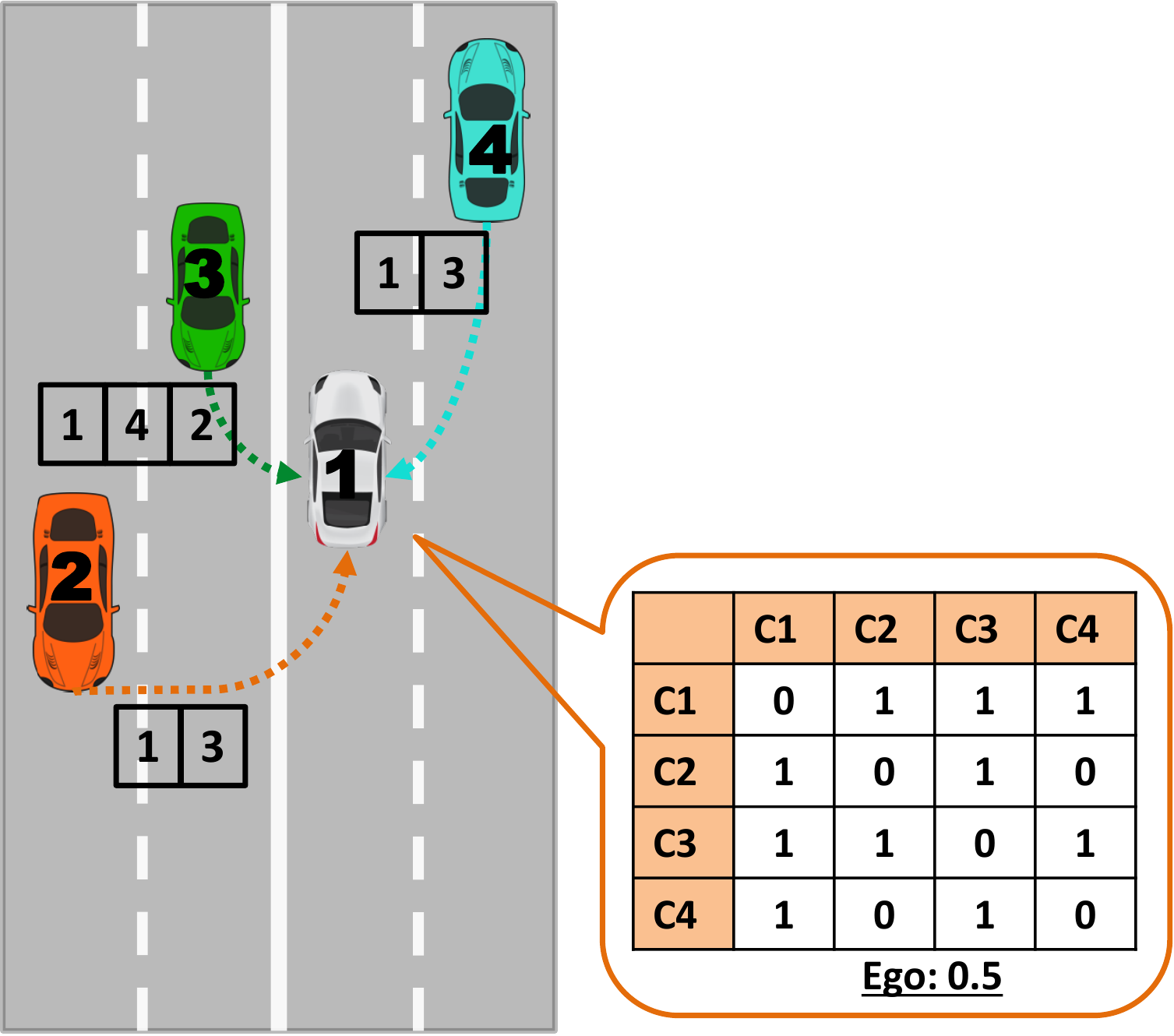}
  \caption{An illustrative example of the ego-network calculation using VANETs.}
\label{egoNetwork_cars}
\end{figure}

The egocentric betweenness is computed by manipulating only the adjacency matrix, i.e., a simple network topology based on one-hop communications. The local information of the network topology is obtained by means of the beaconing approach. Since the vehicle's beacon messages are only useful to one-hop neighbors, these messages are not forwarded. Therefore, the beacon message exchanged between the vehicles is a list of neighbors, as can be observed in Figure~\ref{egoNetwork_cars}. In this illustrative example, the gray vehicle (labeled as 1) receives the list of neighbors of all vehicles currently within its transmission range (vehicles labeled as 2, 3, and 4). Once received, it can construct the adjacency matrix and calculate the egocentric betweenness score, according to Equation~\ref{EBM}. Each one of the vehicles updates the egocentric betweenness score whenever the adjacency matrix is updated.

There is evidence that the betweenness SNA and EBM have the highest correlation in a static network~\cite{marsden2002}. However, new research indicates that the highest correlation can also exist in highly dynamic networks~\cite{akabane2017applying, akabane2018distributed}, such as VANETs.

The other parameter used to compute the $V_{rank}$ is the radio propagation model. This parameter has been added in order that messages travel on a link with less interference. The model applied was the two-ray ground-reflection, which considers both the direct path and the ground reflection path between the transmitter and the receiver antennas~\cite{sommer2012}, as described by Equation~\ref{two-ray}:

\begin{equation}
  \label{two-ray}
  L_{TRI}[dB] = 20log(4\pi \frac{d}{\lambda}|1+\Gamma\exp^\varphi|^{-1})
\end{equation}
where $\lambda$ is the wavelength, $d$ is the Euclidean distance between two vehicles, $\Gamma$ is the reflection coefficient, and $\varphi$ is the phase difference of interfering rays. The procedure for the calculation of the phase difference of interfering rays is given by Equation~\ref{interferencia}:

\begin{equation}
  \label{interferencia}
  \varphi = 2\pi \frac{d_{los}-d_{ref}}{\lambda}, \begin{cases} d_{los}=\sqrt{d^2+(h_t-h_r)^2} \\ 
								d_{ref}=\sqrt{d^2+(h_t+h_r)^2}\end{cases}
\end{equation}
where $d_{los}$ and $d_{ref}$ describe the propagation distance and the reflection distance, respectively. The variables $h_t$ and $h_r$ represent the height at which the antenna of the transmitter and the receiver, respectively, are relative to the ground. In this work, the height applied to both antennas was \unit[149.5]{cm}, which is commonly adopted in the literature~\cite{sommer2012}. Lastly, the reflection coefficient can be calculated by Equation~\ref{reflexao}: 

\begin{equation}
  \label{reflexao}
  \Gamma = \frac{\sin \theta_i-\sqrt{\varepsilon-\cos \theta_i}}{\sin \theta_i+\sqrt{\varepsilon-\cos \theta_i}}, \begin{cases} \sin \theta_i=\frac{h_t+h_r}{d_{ref}} \\ \cos \theta_i=\frac{d}{d_{ref}}\end{cases}
\end{equation}
where $\varepsilon$ represents the reflection from the ground with a value of $1.02$~\cite{sommer2012} and $\theta$ is the angle between the ground and the reflected ray. Finally, the wavelength value was set to \unit[0.051]{m} according to IEEE 802.11p~\cite{ieee2013ieee}.

Using these two parameters, therefore, each vehicle can autonomously compute its own rank score. Once calculated the $V_{rank}$, each one of the vehicles shares it with its one-hop neighbors. 

\subsection{Generation and Distribution of Knowledge}

The dEASY system sends the GNSS data through periodical messages to its one-hop communication neighbors. These messages are sent into the control channel in the form of beacon packets. In addition to the already contained data in the beacon package, two extra information fields were added, namely the current $V_{rank}$ score and the aggregated data. The local knowledge is created by aggregating the beacon data received from the neighborhood. Once the local knowledge has been created, the next step is to share it with the highest-ranked neighbor vehicle. This vehicle was chosen by applying Equation~\ref{EBM}.

The merging of two aggregated values, if needed, may be performed according to the following function: $A_r:=\partial (A_1, A_2)$, where $\partial$ depicts the aggregation function that has two input values ($A_1$ and $A_2$). Therefore, these input values are combined, generating a new aggregated value ($A_r$). As the main goal of this work is the design of a vehicular traffic management system, the aggregation function is given by the Equation~\ref{aggregation_function}:

\begin{equation}
  \label{aggregation_function}
  v^{avg}_{agg_{(i)}} = \frac{v_1 c_1 + v_2 c_2}{c_1 + c_2}
\end{equation}
where $v^{avg}_{agg_{(i)}}$ is the average aggregate speed of a given road $i$. The parameters $v_1 $ and $ v_2 $ are the two input values of the same road, which are going to be aggregated. $c_i$ indicates the amount of data that contributed to the generation of the new aggregated value. Every time the vehicle receives a report concerning a given road $i$, it will smooth the computed aggregate average speed $v^{curr}_{agg_{(i)}}$ using the following Equation~\ref{smooth_aggregation_function}: 

\begin{equation}
  \label{smooth_aggregation_function}
  v^{curr}_{agg_{(i)}} = \sigma v^{old}_{agg_{(i)}} + (1-\sigma) v^{new}_{agg_{(i)}}
\end{equation}

The main goal of the equation is to keep the report updated, where $\sigma$ is the weighting factor. This factor is needed to assign a higher weight to the most current information. In addition to that equation, we applied Equation~\ref{weight_road} to calculate the weight of the road $i$ ($ w_{road_{(i)}} $), which will be used in the classification step. 

\begin{equation}
  \label{weight_road}
  {\displaystyle w_{road_{(i)}} = \frac{v^{avg}_{agg_{(i)}}}{v^{max}_{spe_{(i)}}}}, \begin{cases}
								  w_{road_{(i)}}: \,$weight$\,$of$\,$road$\,i\\
								  v^{avg}_{agg_{(i)}}: \,$aggregate$\,$average$\,$speed$\,$of$\,$road$\,i\\
								  v^{max}_{spe_{(i)}}: \,$maximum$\,$speed$\,$of$\,$road$\,i
								 \end{cases}
\end{equation}

Equation~\ref{weight_road} follows the same criteria of the one used in the Highway Capacity Manual~(HCM)~\cite{HCM2016}. After aggregating all the local data, the vehicle that has the highest $V_{rank}$ score, in that particular moment, classifies the weight of the roads according to the Table~\ref{tab:hcm}. Here again, the levels-of-service and traffic classification were based on the HCM~\cite{HCM2016}.

\begin{table}[!h]
\begin{center}
\small
\caption{Level of service and traffic classification \cite{HCM2016}.}
  \label{tab:hcm}
  \begin{tabular}{ccc}
   \hline
    Level of Service & Traffic Classification & $w_i$ \\
    \hline
    \centering
    \textbf{A}       & Free flow                    & (1.0 $\sim$ 0.9] \\
    \textbf{B}       & Reasonably free flow         & (0.9 $\sim$ 0.7] \\
    \textbf{C}       & Stable flow                  & (0.7 $\sim$ 0.5] \\
    \textbf{D}       & Approaching unstable flow    & (0.5 $\sim$ 0.4] \\
    \textbf{E}       & Unstable flow                & (0.4 $\sim$ 0.33] \\
    \textbf{F}       & Forced or breakdown flow     & (0.33 $\sim$ 0.0] \\
   \hline
 \end{tabular}
\end{center}
\end{table}

\begin{figure}[htp!]
\centering
  \captionsetup{justification=centering}
  \includegraphics[width=0.7\textwidth]{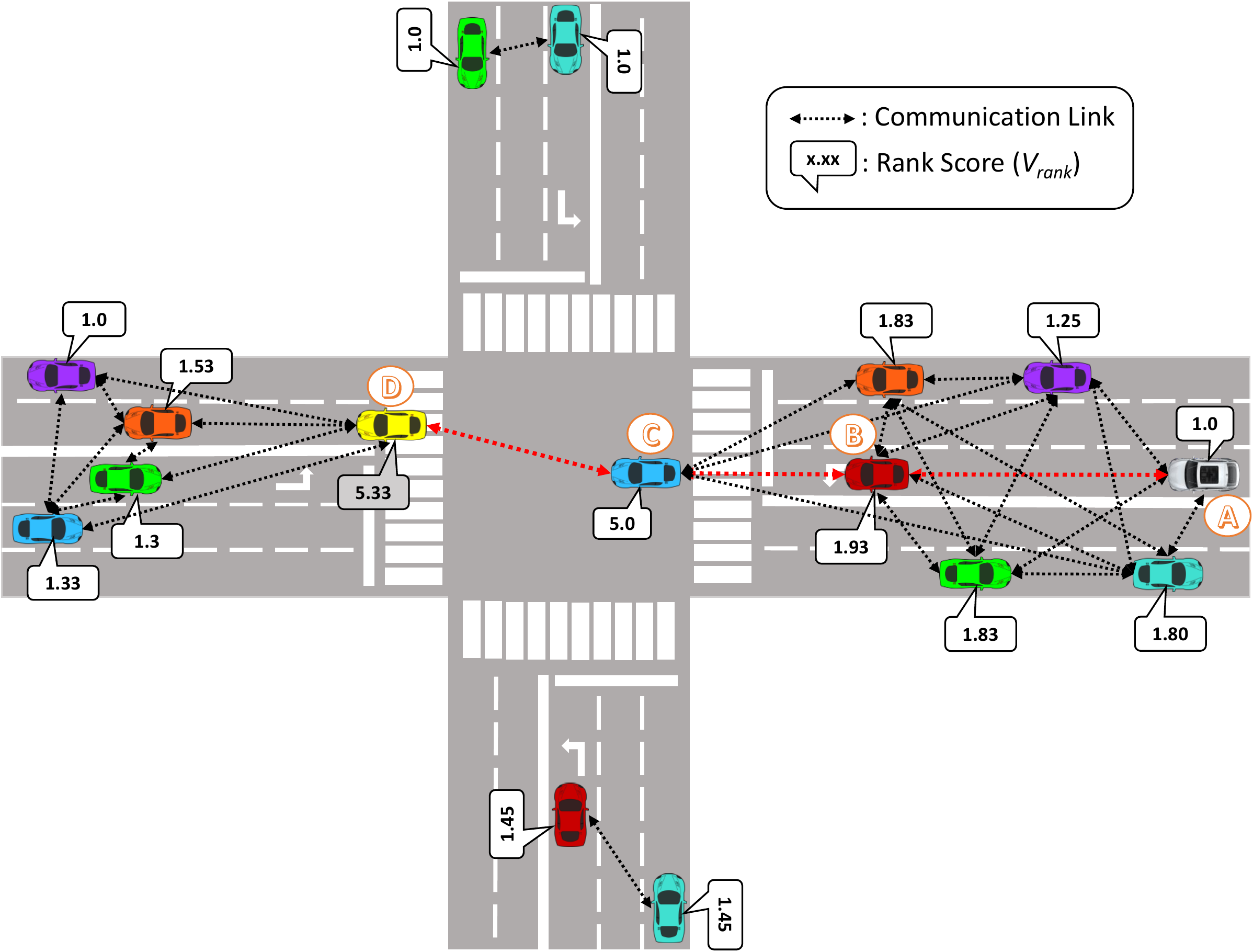}
  \caption{An illustrative example of inter-vehicular communication links.}
\label{example_ego_car}
\end{figure}

Figure~\ref{example_ego_car} describes the inter-vehicular communication links at a given time. In this example, it is assumed that the vehicle labeled as $A$ needs to forward its aggregate local data to the next vehicle within its transmission range. First, it selects the next neighbor vehicle with the highest $V_{rank}$ score. In this case, the vehicle with $1.93$ score~(vehicle $B$). As soon as vehicle $B$ receives the aggregate local data, it aggregates the received data with its own data and forwards it to the next neighbor vehicle with the highest $V_{rank}$ score. This procedure will be repeated until the data reaches the vehicle $D$. It is worth remembering that the vehicles that were not selected will decline the received data. This situation can be observed in Figure~\ref{example_ego_car}, where the vehicle $D$ has the highest $V_{rank}$ score among all the participants, at that moment, thus all the aggregated data will be directed to it. 

While vehicle $D$ has the highest $V_{rank}$ score, it classifies all aggregated data received. If during the classification step an event is identified~(in this case, levels D, E or F of Table~\ref{tab:hcm}), a message (or knowledge) containing the identification of the roads in question is generated. After this step, it initiates the knowledge distribution process in the service channel. The sender's neighboring vehicles that received the knowledge will schedule a retransmission to continue the knowledge distribution process. Every time a vehicle receives a knowledge to be distributed, it checks if it is within the zone of preference (ZoP)~\cite{akabane2015}, if that holds true, it transmits first. Due to the broadcast suppression mechanism implemented, as soon as the neighboring vehicles outside the zone of preference receive the same scheduled knowledge, they cancel the retransmission. This allows avoiding the traffic of redundant knowledge and also decreases network overhead.

\subsection{Consumption of the knowledge} \label{consumption_knowledge_layer}

The knowledge generated as a notification message about congestion is propagated through the VANET. Thereby, vehicles approaching the congestion spot can receive and use this knowledge to compute an alternative route in a timely manner. 

\begin{algorithm}
\caption{Knowledge consumption and broadcast procedure}
 \SetKwInOut{Input}{inputs}
 \SetKwInOut{Output}{output}
 \label{consumption_knowledge}
   \Input{Notification message ($m_{cong}$), containing the identification of the congested roads $and$ the route information message ($m_{rou}$)}
   \If{$receive\,m_{cong}$}{
    \If{$my\,route\,crosses\,congestion\,AND\,can\,be\,avoided$}{
      $waiting\_time = ComputeWaitingTime()$\;
      \While{$waiting\_time\,AND\,receive\,m_{rou}$}{
	$CollectRouteInformation(m_{rou})$\;
      }
      $ComputesWeightPopularityRoadSegment()$\;
      $np = EntropyBasedRoutePlanning()$\;
      $BroadcastNewPath(np)$\;
    }
    \Else{
      $DiscardsMessage(m_{cong});$
    }
   }
\end{algorithm}

When a vehicle receives the knowledge, it triggers the procedure described in Algorithm~\ref{consumption_knowledge}. The vehicle first checks if the current route is traveling through the congestion spot as well as the possibility of avoiding it (Line 2). The received knowledge will be used in the alternative route computation step together with the alternative route information received from surrounding vehicles. This means that the vehicles compute an alternative route based on an altruistic routing decision when congestion is identified. Before calculating an alternative route, the vehicle waits for a short time~(Line 3). The waiting time is directly proportional to the distance between the congestion spot and the vehicle's current position. The purpose of this waiting time is to collect alternative routes from surrounding vehicles. In addition to that, the idea here is to prioritize vehicles closest to the congestion spot to compute an alternative route first. Note that the first vehicle computes an alternative route without considering the others. From the second vehicle onwards, the neighborhood information route is taken into account for the calculation~(Lines 4 and 5). In order to do that, Definition 1 shows how the weight of the popularity of the road segment is calculated. The idea of this measure is to verify to which roads the vehicles are being moved to. As soon as the waiting time expires, the weight of the road segment is calculated according to Definition 1 (as depicted in Line 6 in Algorithm~\ref{consumption_knowledge}). After this step is completed, the route planning~(Line 7) is computed through the use of the Entropy-based Shortest-Path, as described in Definition 2. The alternative route with the lowest entropy score will be chosen. In other words, the priority is to choose an alternative route with a low popularity score, therefore, avoiding a potential secondary congestion. Lastly, in Line 8, the alternative route selected will be broadcast to neighbors.

\vspace{0.1cm}
\noindent
\textit{\textbf{Definition 1 - Weight of the Popularity of the Road Segment} is described as $w_{rs_{(i)}} = n_{(i)} \times w_{(i)}$, where $n_{(i)}$ describes the total number of vehicles that will pass through road segment $i$, and $w_{(i)}$ depicts a weight associated to it. The calculation of the weight is given by $w_{(i)} = \frac{len_{(i)}}{lane_{(i)}}$, where $len_{(i)}$ and $lane_{(i)}$, represent the length and the number of lanes of the $r_i$, respectively.}

\vspace{0.2cm}
\noindent
\textit{\textbf{Definition 2 - Entropy-based Shortest-Path}: let $R = \{r_1, ..., r_n\}$ be the path computed by the vehicle and let ($w_{rs_1}, ..., w_{rs_n}$) be the set of the weight score associated with each road segment of $R$. Thereby, the entropy measure of the vehicle's path $v$, is defined as $H_{p_v} = \sum_{i=1}^{n} \frac{w_{rs_{(i)}}}{Q} \times \log(\frac{Q}{w_{rs_{(i)}}})$, where $Q = \sum_{i=1}^{n}w_{rs_{(i)}}$.}

\begin{figure*}[htp!]
\centering
  \captionsetup{justification=centering}
  \includegraphics[width=1.05\textwidth]{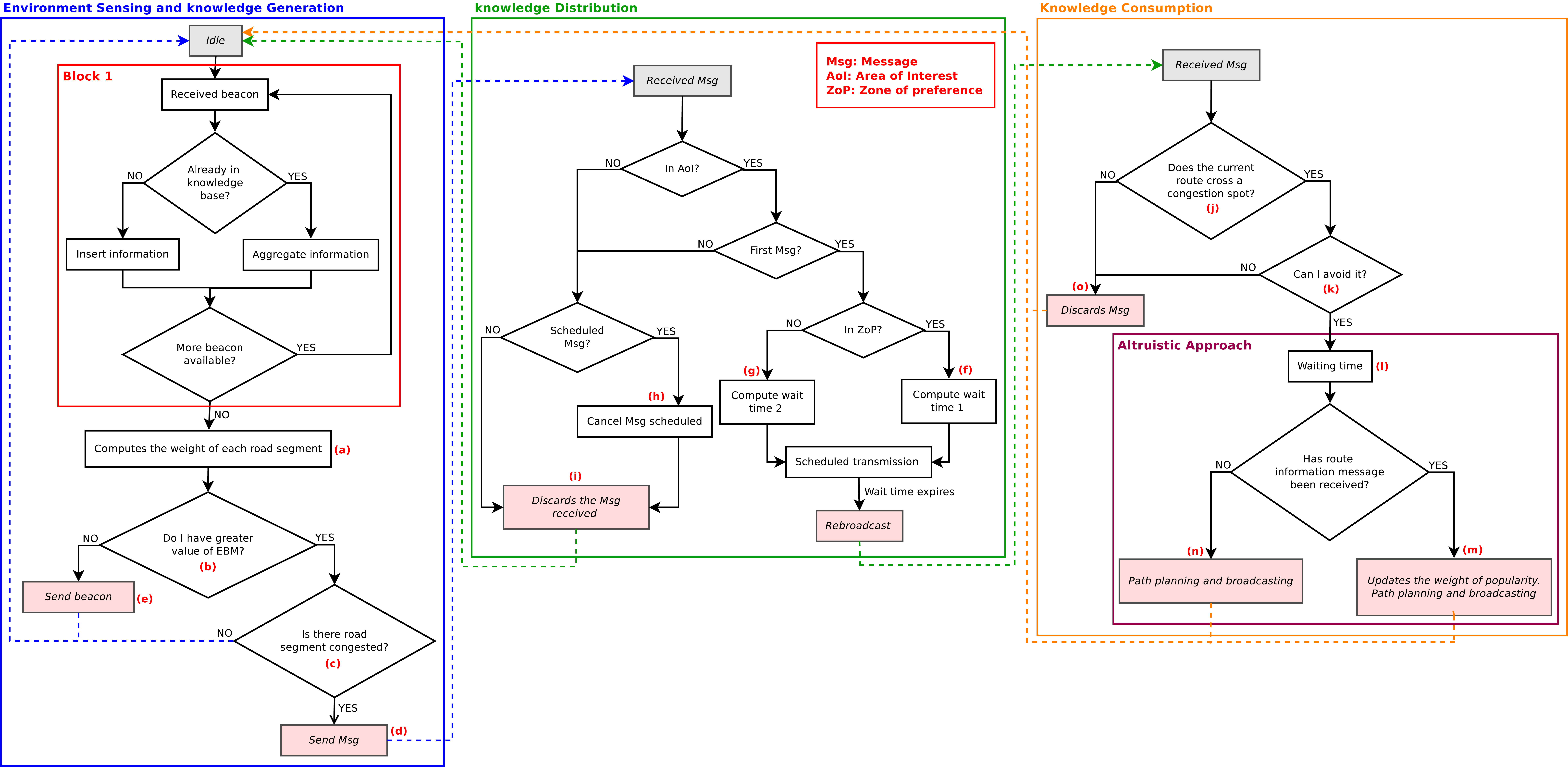}
  \caption{The dEASY's operation flowchart.}
\label{system_flow}
\end{figure*}

\vspace{0.1cm}
Figure~\ref{system_flow} shows a flowchart illustrating the steps followed by the dEASY's system to contemplate the three-layer architecture. When dEASY receives the local information, it either inserts or aggregates, according to Equation~\ref{smooth_aggregation_function}~(Block 1). After this step, it calculates the weight of the roads according to Equation~\ref{weight_road}~(Label~(a)). In addition, if the vehicle has the highest EBM value~(Label~(b)), it also classifies the weight of the roads according to Table~\ref{tab:hcm}~(Label~(c)). In this process, if the selected vehicle finds out that there is a congested traffic flow, the knowledge about this event is generated and distributed in the network~(Label~(d)). On the other hand, if the vehicle does not have the highest EBM value, it selects the next most relevant vehicle, following the procedure presented at Subsection~\ref{egocentric}, and sends the aggregated local information to it~(Label~(e)).

To avoid the well-known broadcast storm problem, a broadcast suppression mechanism that uses ZoP concept~\cite{akabane2015} was applied. Only vehicles that are located inside the ZoP will continue the knowledge forwarding process. The ZoP mechanism works as follows: each vehicle within the area of interest computes its own waiting time as soon as it receives the first knowledge. However, vehicles that are within the ZoP have a smaller waiting time~(Label~(f)) than other vehicles outside of it~(Label~(g)). Thereby, vehicles that are on the outside will receive redundant messages from the vehicles located within the ZoP. These vehicles will cancel the scheduled transmission~(Label~(h)) and the message received will be dropped~(Label~(i)). Vehicles outside the ZoP will be requested only when there are no other vehicles within the ZoP.

Every vehicle that receives the knowledge about a congestion spot will first verify if the current route is moving towards the congestion~(Label~(j)) as well as if it can be avoided~(Label~(k)). If so, it waits a period of time~(Label~(l)) before computing an alternative route. This time interval is needed for the altruistic approach since the calculation of an alternative route also depends on the neighborhood routes. After the waiting time expires, if the vehicle has received information about alternative routes from the neighborhood, these will be taken into account for the calculation of its own alternative route~(Label~(m)). This procedure is computed as described in Entropy-based Shortest Path~(Definition 2). Otherwise, the vehicles compute the best alternative route~(Label~(n)). Lastly, if the vehicle does not pass through the congested area, it simply discards the knowledge~(Label~(o)).

\section{Simulation Setup}

A set of experiments were conducted to assess the proposed system using the vehicular network simulator framework, known as Veins~\cite{SOMMER2011}. It integrates the OMNeT++ network simulator~\cite{VARGA2001} with the SUMO road traffic simulator~\cite{KRAJZEWICZETAL2002}. The Physical~(PHY) and Medium Acess Control~(MAC) layers are implemented in Veins and based on the IEEE 802.11p~(WAVE) standard. 

The bitrate was set to~\unit[6]{Mbps} in the MAC layer, the transmission power used was \unit[20]{mW} under a two-ray ground propagation model~\cite{sommer2012}, and the receiver sensitivity was set to \unit[-82]{dBm}. In addition to that, the movement of the vehicles is based on the~\textit{Krauss}' car-following model~\cite{krauss1998}.

\begin{figure}[htp!]
\centering
  \captionsetup{justification=centering}
  \includegraphics[width=0.5\textwidth]{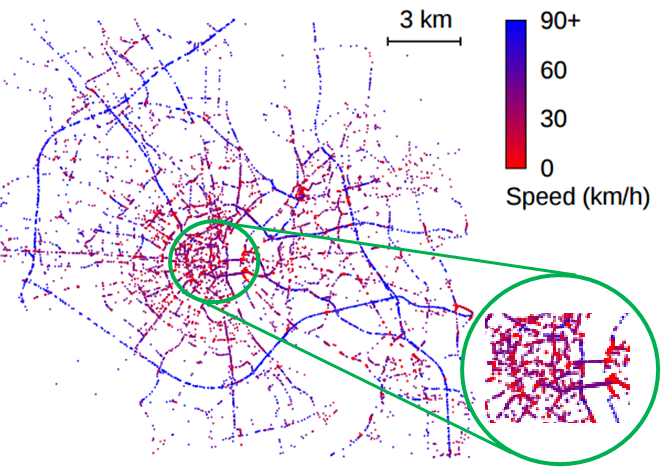}
  \caption{Urban area of the city of K\"oln, Germany - adapted \cite{uppoor2012large}.}
\label{TAPASCologne}
\end{figure}

In order to evaluate a vehicular traffic management system, a real-world scenario of urban mobility should be considered. Based on this, the TAPASCologne project\footnote{http://sumo.dlr.de/wiki/Data/Scenarios/TAPASCologne}, of the Institute of Transportation Systems at the German Aerospace Center (ITS-DLR), was applied in our simulation experiments as the evaluation scenario. This project aims to reproduce the vehicle traffic in a large-scale scenario of the city of K\"oln, Germany, with the highest possible level of realism. This dataset includes traffic demand from \unit[6:00]{am} to \unit[8:00]{am}, with more than 250,000 vehicles routes. However, only a central submap of the city of K\"oln was chosen for our simulation experiments because it displays a higher incidence of traffic congestion, as highlighted in Figure~\ref{TAPASCologne}. Moreover, only routes of vehicles that pass through the submap were selected. The new dataset was also divided into five different penetration rates - \unit[20]{\%}, \unit[40]{\%}, \unit[60]{\%}, \unit[80]{\%}, and \unit[100]{\%}. This means that, of the total of the new dataset, only \unit[20]{\%} of the vehicles are inserted in the scenario for our simulation experiments, and so on. All the experimental results of this work were conducted with a confidence interval of \unit[95]{\%}. Table~\ref{table:parameters} summarizes the simulation parameter settings.

\begin{table}[!h]
\centering
\caption{Simulation parameters settings.}\label{table:parameters}
\small
  \begin{tabular}{l l}
  \hline 
  \multicolumn{1}{l}{\textbf{Parameter}} 	& \textbf{Value} \\ \hline \hline \\[-0.3cm]
  MAC layer				 	& IEEE 802.11p PHY \\
  Bandwidth				 	& \unit[10]{MHz} \\
  NIC Bitrate                       		& \unit[6]{Mbps} \\
  NIC TX power            			& \unit[20]{mW} \\
  NIC Sensitivity				& \unit[-82]{dBm} \\
  Transmission range            		& \unit[287]{m} \\
  Beacon transmission rate 			& \unit[1]{Hz} \\
  Alpha ($\alpha$)                     & \unit[0.5]{} \\
  Confidence interval           		& \unit[95]{\%} \\
  \hline
  \end{tabular}
\end{table}

Twelve metrics were used to evaluate the performance of the dEASY system. These metrics were divided into three perspectives~(or assessments), which are described in detail below. Each perspective corresponds to a layer of the dEASY's architecture.

\begin{enumerate}
 \item Control channel assessment (\textit{Environment Sensing and Vehicle Ranking Layer}).
  \begin{itemize}
   \item \textbf{Channel busy ratio}: indicates the interference level and is estimated as the fraction of the time in which the channel is identified as busy, due to packet collisions or successful transmission to the total time;
    \item \textbf{Total transmitted beacon}: shows the number of beacon packets transmitted in the network by all the vehicles during the simulation run;
    \item \textbf{Beacon transmitted per vehicle}: gives the number of beacon packets transmitted per each vehicle during the simulation run.
  \end{itemize}
  \item Scalability assessment (\textit{Knowledge Generation and Distribution Layer}).
  \begin{itemize}
    \item \textbf{Coverage}: indicates the percentage of messages delivered;
    \item \textbf{Overhead}: measures the total amount of transmitted messages by the vehicles;
    \item \textbf{Delay}: demonstrates the time spent to deliver the messages to the vehicles;
    \item \textbf{Collision}: shows the total number of packet collisions during message transmission.
  \end{itemize}
   \item Traffic management assessment (\textit{Knowledge Consumption Layer}).
  \begin{itemize}
    \item \textbf{Travel distance}: shows the average distance traveled by all vehicles;
    \item \textbf{Travel time}: indicates the average travel time in relation to all vehicles;
    \item \textbf{Congestion time loss}: describes the average time spent on congestion;
    \item \textbf{CO$_2$ emission}: gives the average CO$_2$ emission of all vehicles;
    \item \textbf{Planning time index}: measures the total time needed to plan for an on-time arrival 95\% of the time~\cite{lyman2008}. It is calculated as the rate of the 95th percentile travel time to the free-flow travel time.
  \end{itemize}
\end{enumerate}

\section{Simulation Results} \label{simulation_results}

The aim of this section is to assess the performance of dEASY in comparison to the EcoTrec and DIVERT systems. In addition, the original vehicular mobility trace of the K\"oln (OVMT) is going to be used as a baseline since it does not apply any vehicle routing mechanism. For a better presentation, the results were divided into three subsections~(\ref{control_channel_assessment}, \ref{scalability_assessment}, and \ref{traffic_management_assessment}) according to the dEASY's architecture layers.

\subsection{Control channel assessment}\label{control_channel_assessment}

\begin{figure*}[!h]
	\centering
	\captionsetup{justification=centering}
	\subfigure[Channel busy ratio.]{
	\label{channel_busy}
	\includegraphics[width=.47\textwidth]{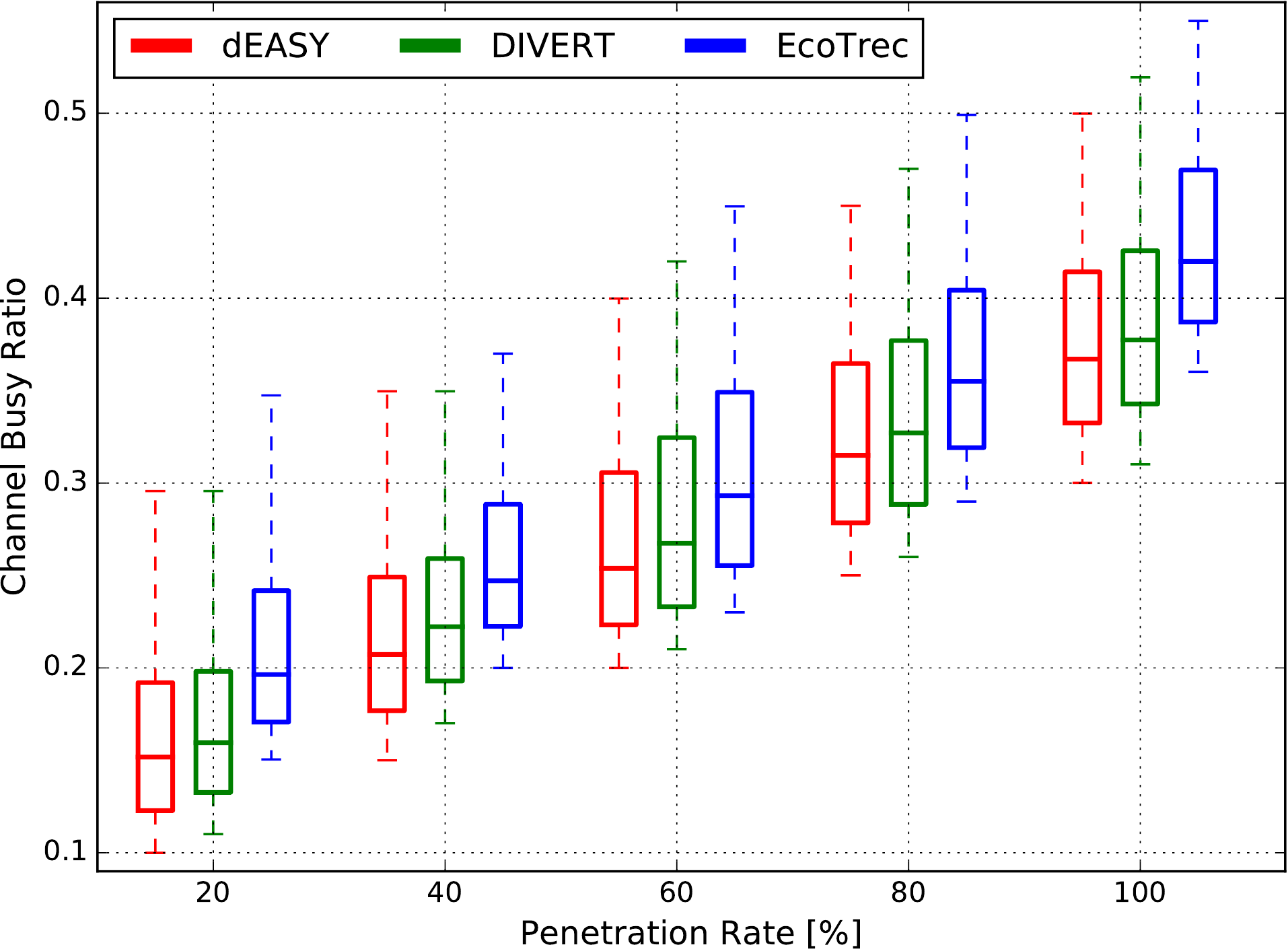}
	}	
	\subfigure[Beacon packets transmitted per vehicle.]{
	\label{beacons_vehicle}
	\includegraphics[width=.47\textwidth]{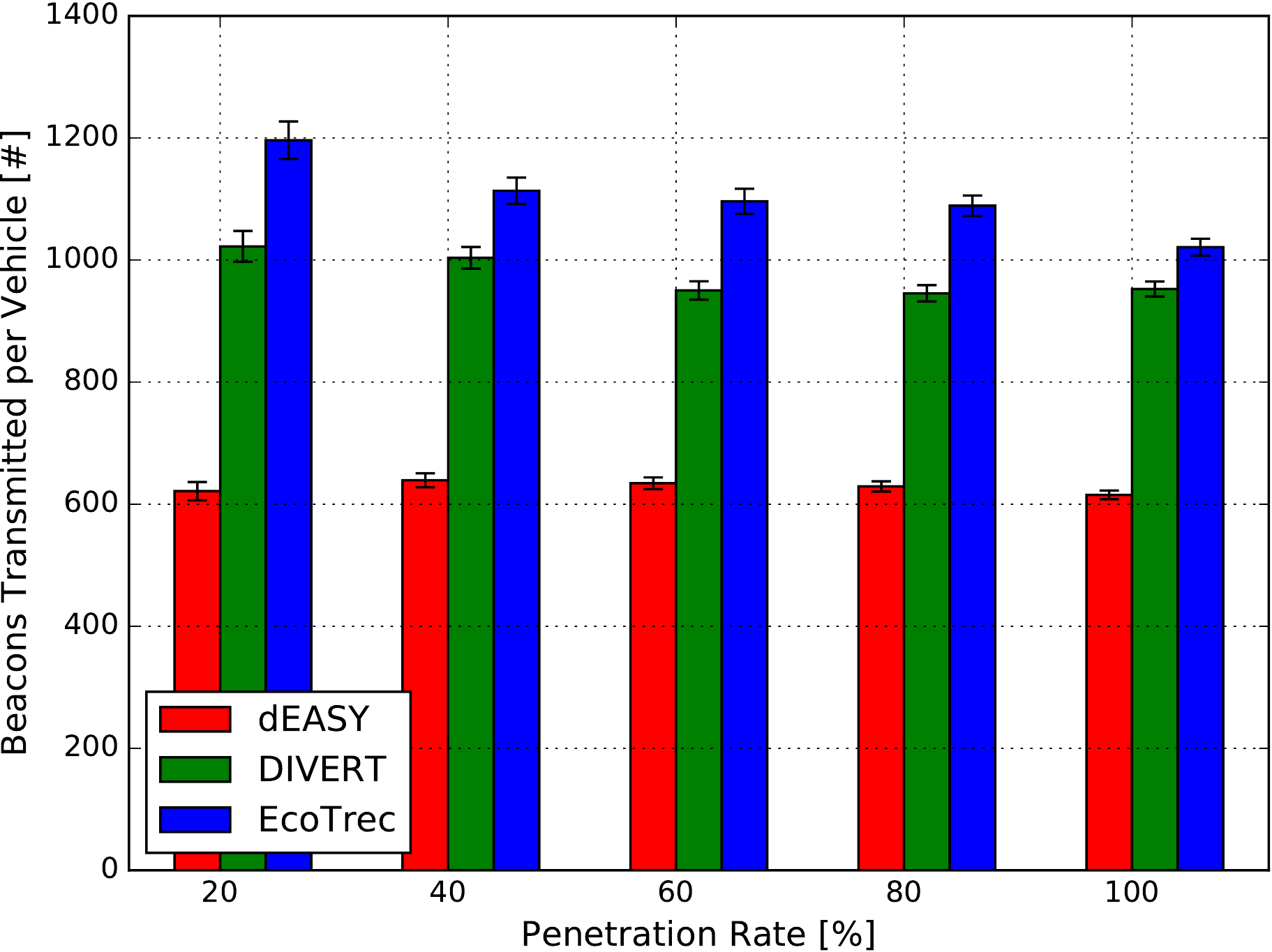}
	}
	\subfigure[Total beacon packets transmitted.]{
	\label{total_beacons}
	\includegraphics[width=.47\textwidth]{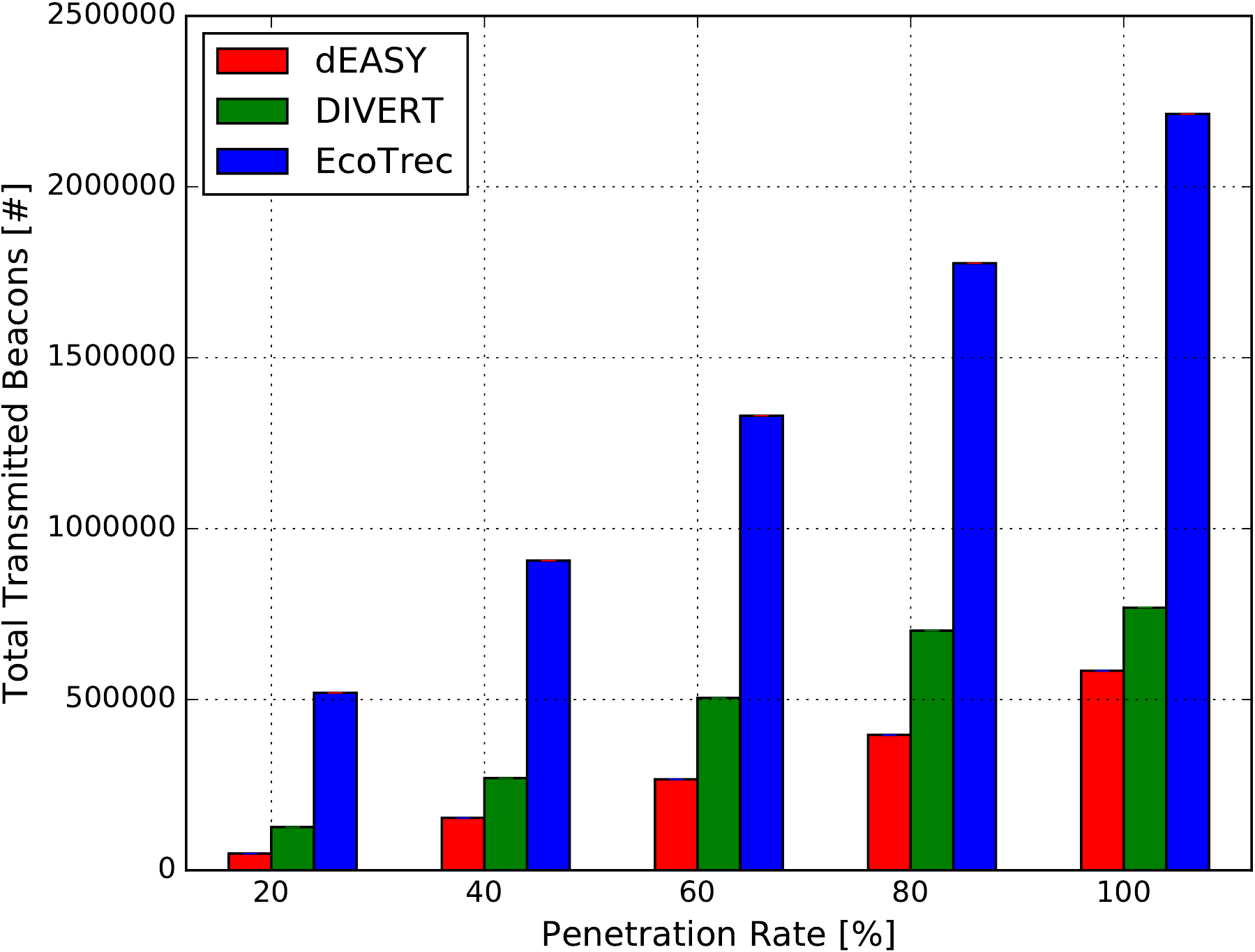}
	}
	\caption{The results of the control channel assessment.}
	\label{control_channel}
\end{figure*}

The assessment of the control channel is important since all the solutions analyzed apply the beaconing approach to share context-awareness information. In the experiments, the beacon transmission rate of 1Hz was set to all systems~\cite{doolan2017ecotrec, pan2017divert}, including dEASY.

Figure~\ref{control_channel} depicts the performance results of the control channel at all the penetration rate possibilities. In addition, Figure~\ref{channel_busy} shows the average channel busy ratio for each penetration rate. As expected, the channel busy ratio increases with the penetration rate. This is due to the fact that the number of vehicles in the neighborhood increases, thereby, raising the channel busy state. It is important to notice that dEASY has the lowest average channel busy ratio for all penetration rates, when compared to DIVERT and EcoTrec. This can be attributed to the fact that it has a better vehicular traffic management, as it will be explained later. In other words, dEASY distributes the vehicles in a way that makes the most of the availability of roads. Consequently, the homogeneous distribution of vehicles on the roads reduces the demand on the control channel.

The adoption of good traffic management practices can sharply reduce the vehicle travel time. A direct consequence of this is the reduction of the number of beacon packets transmitted by vehicles. Figure~\ref{beacons_vehicle} shows a microscopic view of the average number of the beacon packets transmitted by each vehicle. Through this figure, it can be observed that dEASY has the least number of beacon packets transmitted when compared to the DIVERT and EcoTrec. Based on this metric it can be assumed that dEASY has the best traffic management compared to its counterparts. It is worth noting that this assumption will be confirmed later. A macroscopic view is depicted in the Figure~\ref{total_beacons}, where it is displayed the total number of beacon packets transmitted during the entire simulation. As expected, dEASY has the smallest total number of packets transmitted, followed by DIVERT and EcoTrec.

The WAVE standard has a single control channel and for better use of this channel one or both practices can be applied: \textit{(i)} a low rate of packet transmission and/or \textit{(i)} good traffic management practices. In our case, both practices were taken into account. The main lesson assimilated from these results is that the beacon transmission rate of \unit[1]{Hz}, may be considered suitable for the scenario evaluated together with the adopted mobility model for the dEASY system. Because the channel busy rate was around \unit[36]{\%}, on average, at the maximum analyzed penetration rate, as shown in Figure \ref{channel_busy}.

\subsection{Scalability assessment}\label{scalability_assessment}

\begin{figure*}[!h]
	\centering
	\captionsetup{justification=centering}	
	\subfigure[Total of transmitted messages.]{
	\label{fig:transmitted}
	\includegraphics[width=.47\textwidth]{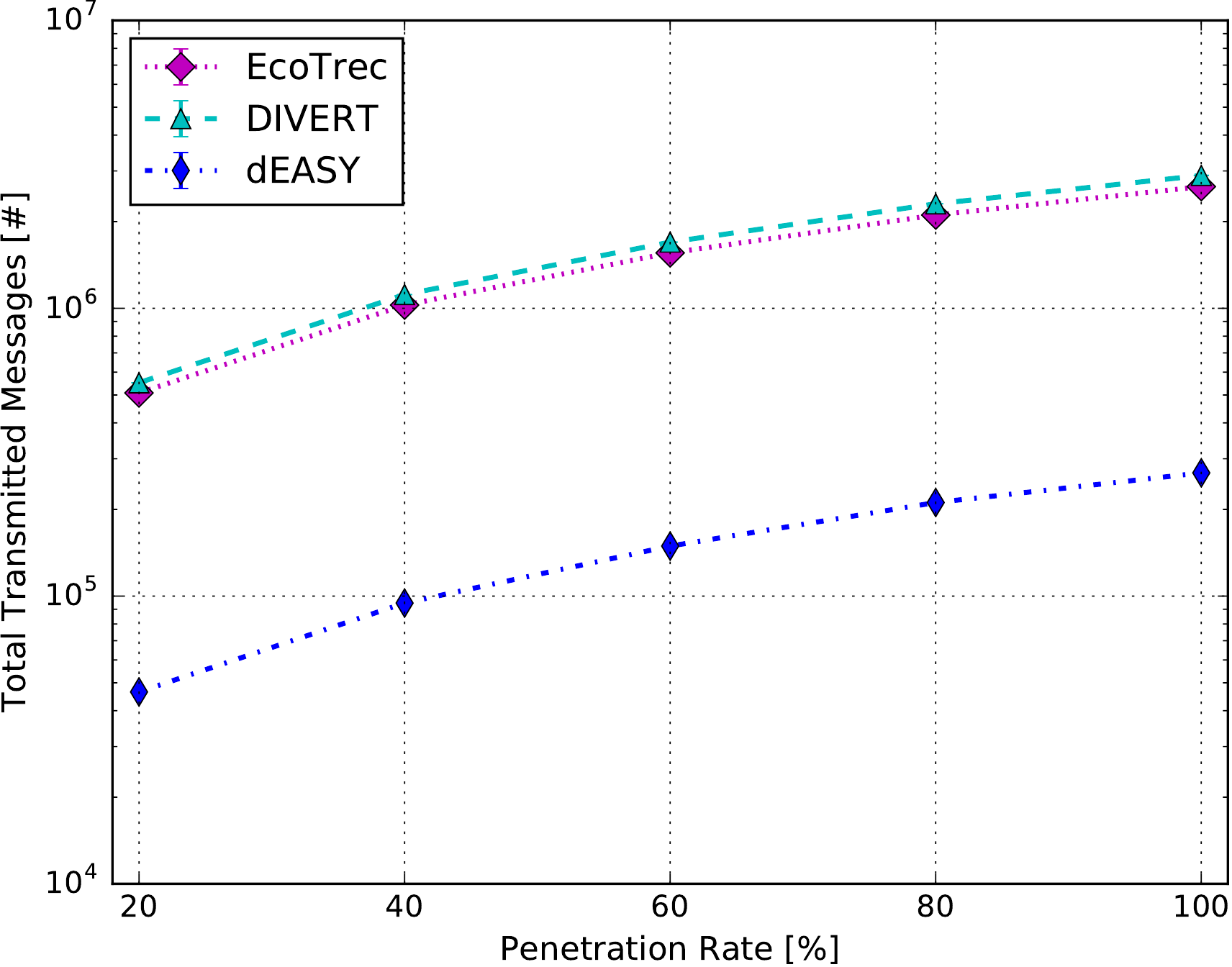}
	}
	\subfigure[Collision.]{
	\label{fig:collision}
	\includegraphics[width=.485\textwidth]{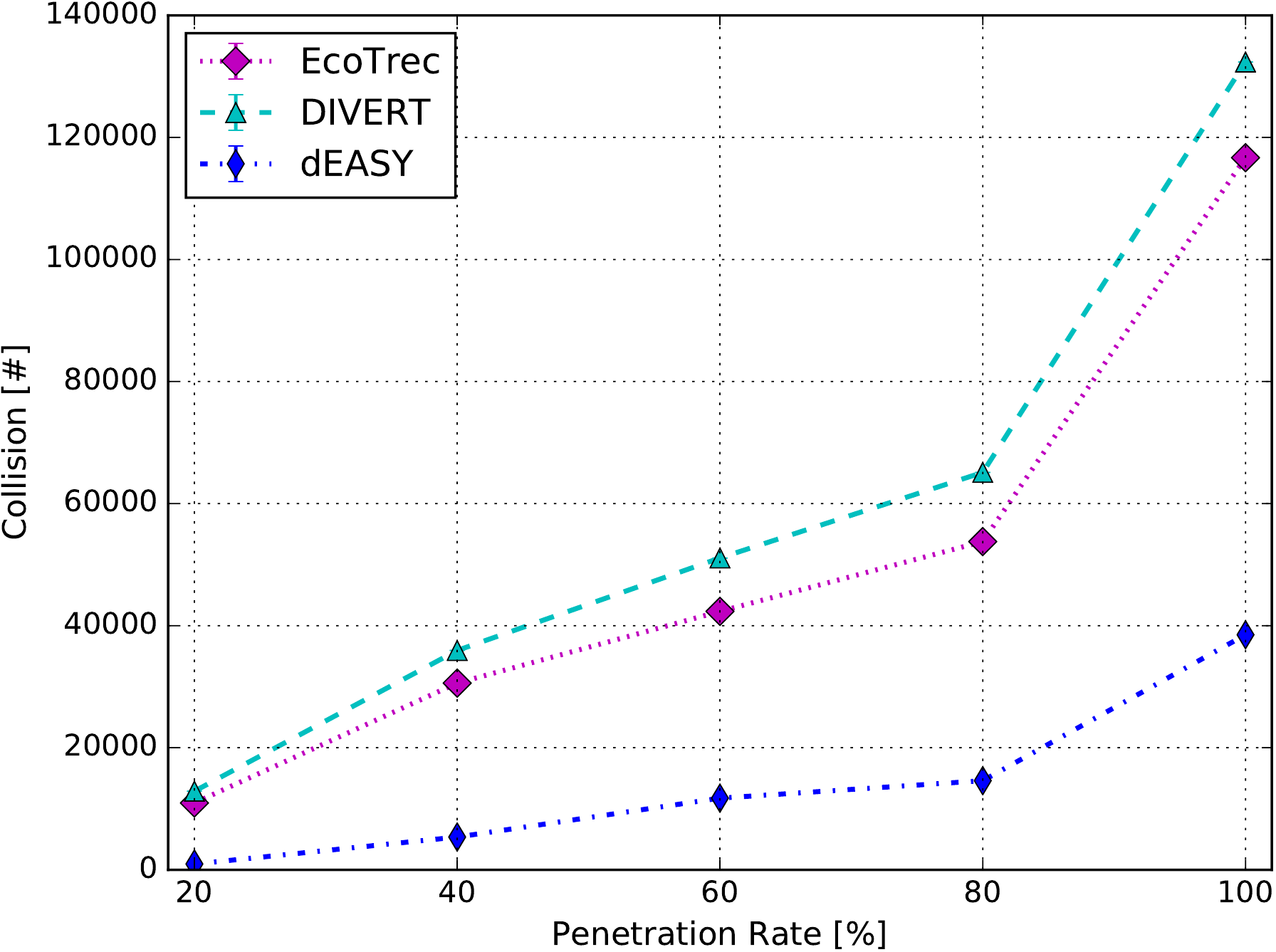}
	}
	\subfigure[Delay.]{
	\label{fig:delay}
	\includegraphics[width=.47\textwidth]{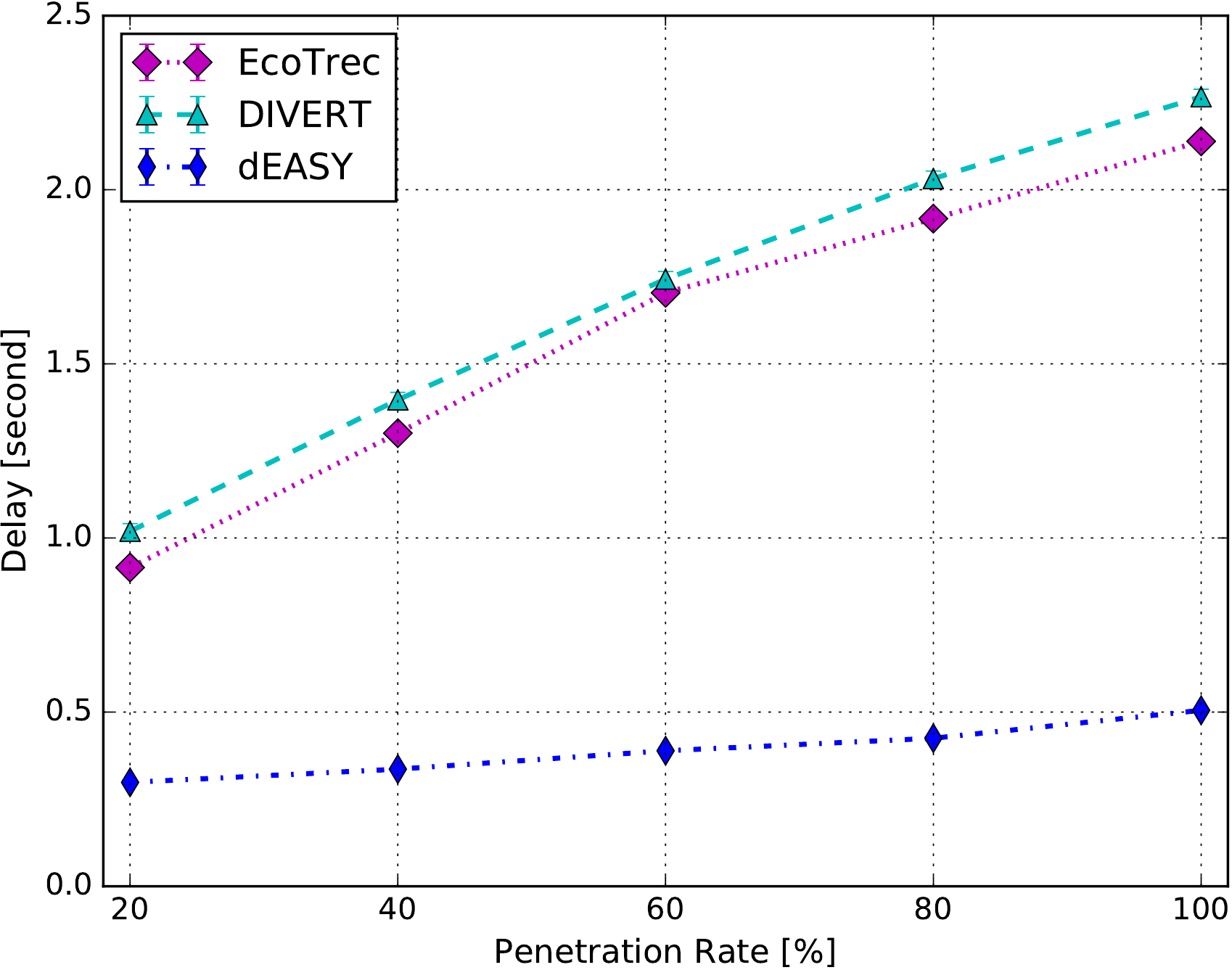}
	}
	\subfigure[Coverage.]{
	\label{fig:coverage}
	\includegraphics[width=.47\textwidth]{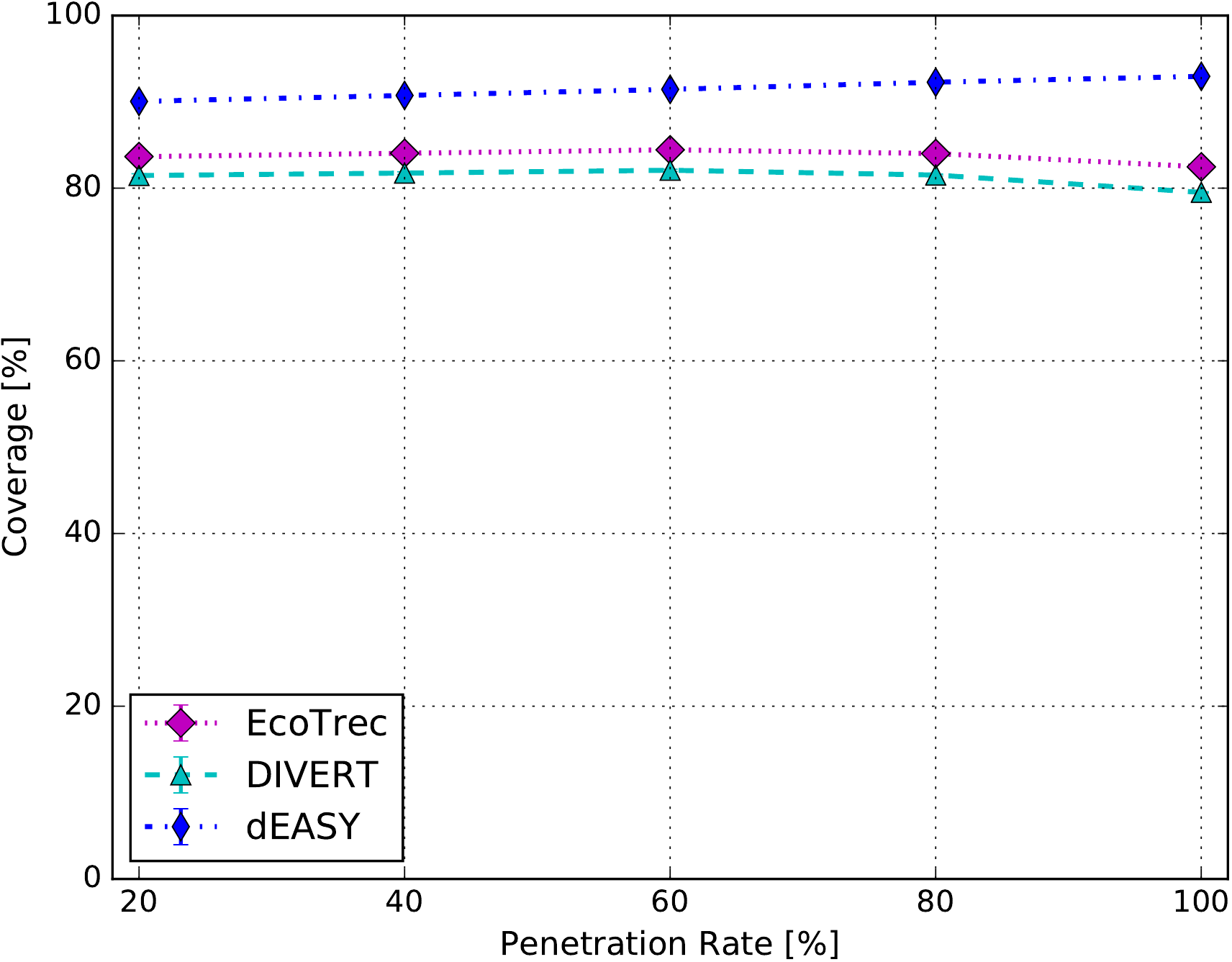}
	}
	\caption{The results of the scalability assessment.}
	\label{scalability}
\end{figure*}

This subsection analyzes the scalability of dEASY against the DIVERT and EcoTrec systems regarding coverage, overhead, delay, and collision metrics. The results are shown in Figure~\ref{scalability}. More specifically, Figure~\ref{fig:transmitted} presents the performance results of all systems investigated according to the overhead metric. 
As expected, the higher the penetration rate, the greater the network overhead for all systems analyzed.
It is known that both EcoTrec and DIVERT systems apply neither the vehicle selection mechanism, to perform the tasks of information aggregation and knowledge generation, nor the broadcast suppression mechanism during the knowledge distribution process. The absence of both mechanisms results in the largest number of messages transmitted between vehicles and vehicle-to-infrastructure. It is also possible to observe that DIVERT curve in the graph is slightly above than EcoTrec. This is related to the fact that DIVERT applies the altruistic routing mechanism in the selection of an alternative route. It is worth noting that the use of this mechanism will help in the choice of alternative routes and the advantage of its use will be discussed later in this paper. The dEASY system applies the vehicle ranking mechanism to select the most appropriate one to perform the information aggregation and knowledge generation. In addition, another mechanism was included to deal with the broadcast storm problem during the knowledge distribution process. The combination of these mechanisms enables dEASY to outperform all the other analyzed systems. Thereby, the proposed system is able to dramatically reduce the total number of messages transmitted in the network, on average, more than \unit[78]{\%} and \unit[75]{\%} compared to DIVERT and EcoTrec, respectively.

Figure~\ref{fig:collision} shows packet collision as the function of the penetration rate. As expected, both EcoTrec and DIVERT systems have a greater number of packet collisions, as this is directly related to the number of packets transmitted on the network. It is known that both systems have greater network overhead compared to dEASY, as shown by the abovementioned figure. On the other hand, the direct consequence of the reduction of the network overhead is the lower packet collision rate and this can be clearly observed in the dEASY system. It achieves an average reduction of up to \unit[64]{\%} and \unit[69]{\%} compared to the EcoTrec and DIVERT systems, respectively.

Another metric evaluated is the transmission delay as the function of the penetration rate, Figure~\ref{fig:delay}. In an infrastructure approach, all vehicles in the scenario constantly take part in contending to access the service channel to send their information. Thereby the packet collisions easily occur mainly as the penetration rate increases, Figure~\ref{fig:collision}, which will result in more data retransmissions and incur an extra delay. On the other hand, the dEASY system employs an infrastructure-less approach, and for this two mechanisms are applied, the vehicle ranking and the broadcast suppression, to decrease transmission and collision rates as well as delay. By comparison, the average transmission delay of dEASY, EcoTrec and DIVERT systems is 0.8, 1.6, and 1.68, respectively. This means that the dEASY system has an average delay reduction of up \unit[50]{\%} and \unit[52]{\%} compared to the EcoTrec and DIVERT systems, respectively. Thus, the infrastructure approach achieves the worst performance in relation to the delay metric, when compared to our proposed system.

Figure~\ref{fig:coverage} illustrates the impact of the penetration rate on the coverage ratio. EcoTrec has a slightly better coverage than DIVERT due to the fact that it presents lower network overhead (Figures~\ref{fig:transmitted} and~\ref{fig:collision}), when compared to its opponent. Another observation is that for high penetration rates (\unit[80]{\%} and \unit[100]{\%}) both systems have a reduction in coverage due to the network overhead. In other words, as the network becomes dense, the service channel competition will lead to more packet collisions, easily resulting in packet transmission failure. On the other hand, as the dEASY system has the least network overhead, the knowledge dEASY generated can reach a larger number of vehicles at all the analyzed penetration rates, reaching, on average, \unit[90]{\%} of the vehicles. This result represents an increase of \unit[11]{\%} and \unit[10]{\%} in comparison to DIVERT and EcoTrec, respectively.

Two main lessons can be drawn from the above results. First, it is clear the advantage of using the proposed vehicle ranking ($V_{rank}$) to select the most relevant vehicle to perform information aggregation and knowledge generation. Moreover, from the analysis of the results, it is possible to notice that the combined use of $V_{rank}$ and the zone of preference enables the construction of a distributed system which can achieve high levels of scalability. Second, the $V_{rank}$ is a viable option for a vehicle selection mechanism in highly dynamic networks.

\subsection{Traffic management assessment}\label{traffic_management_assessment}

\begin{figure*}[!h]
	\centering
	\captionsetup{justification=centering}
	\subfigure[Travel time.]{
	\label{fig:travel_time}
	\includegraphics[width=.47\textwidth]{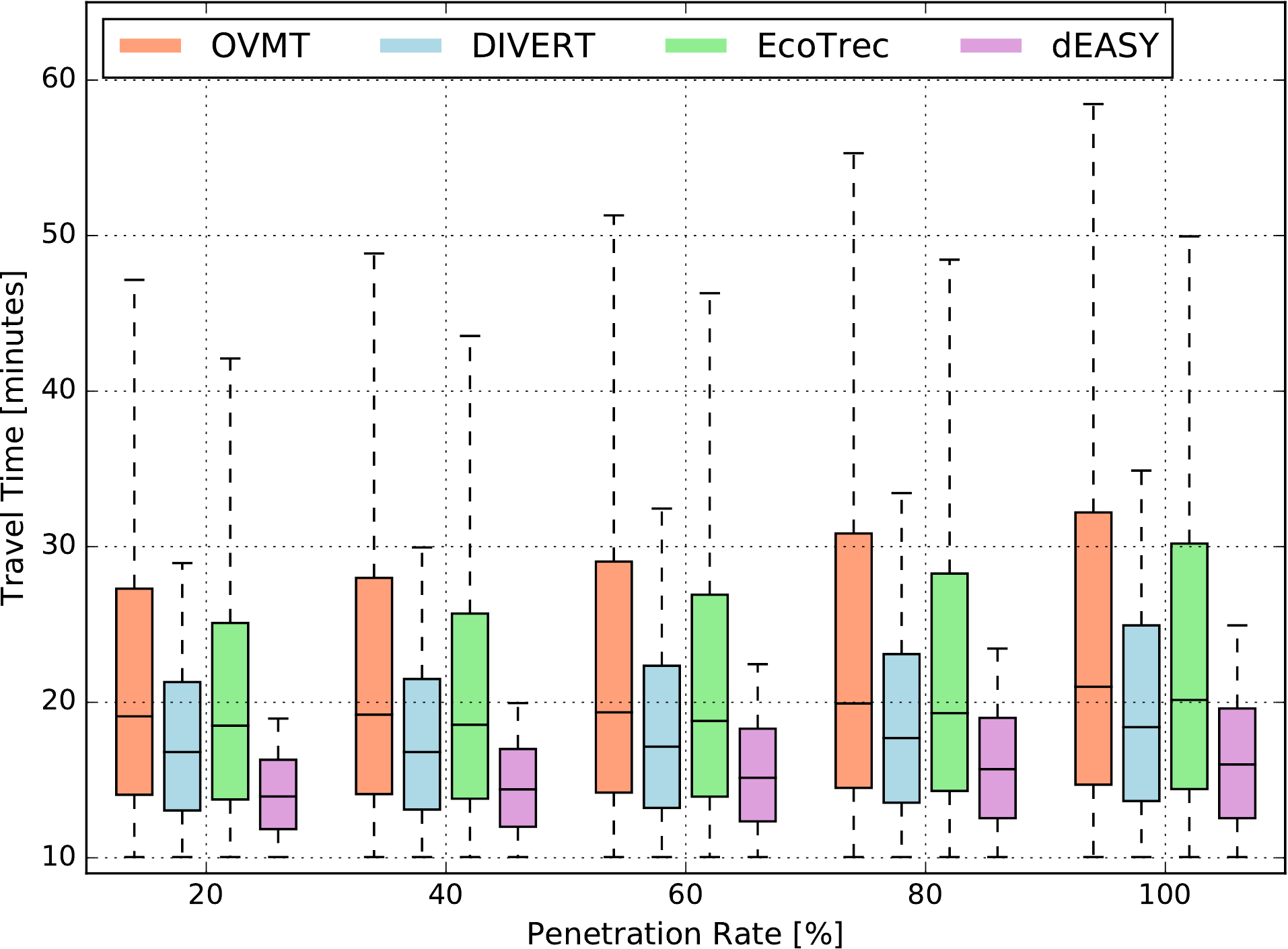}
	}	
	\subfigure[Travel distance.]{
	\label{fig:travel_distance}
	\includegraphics[width=.47\textwidth]{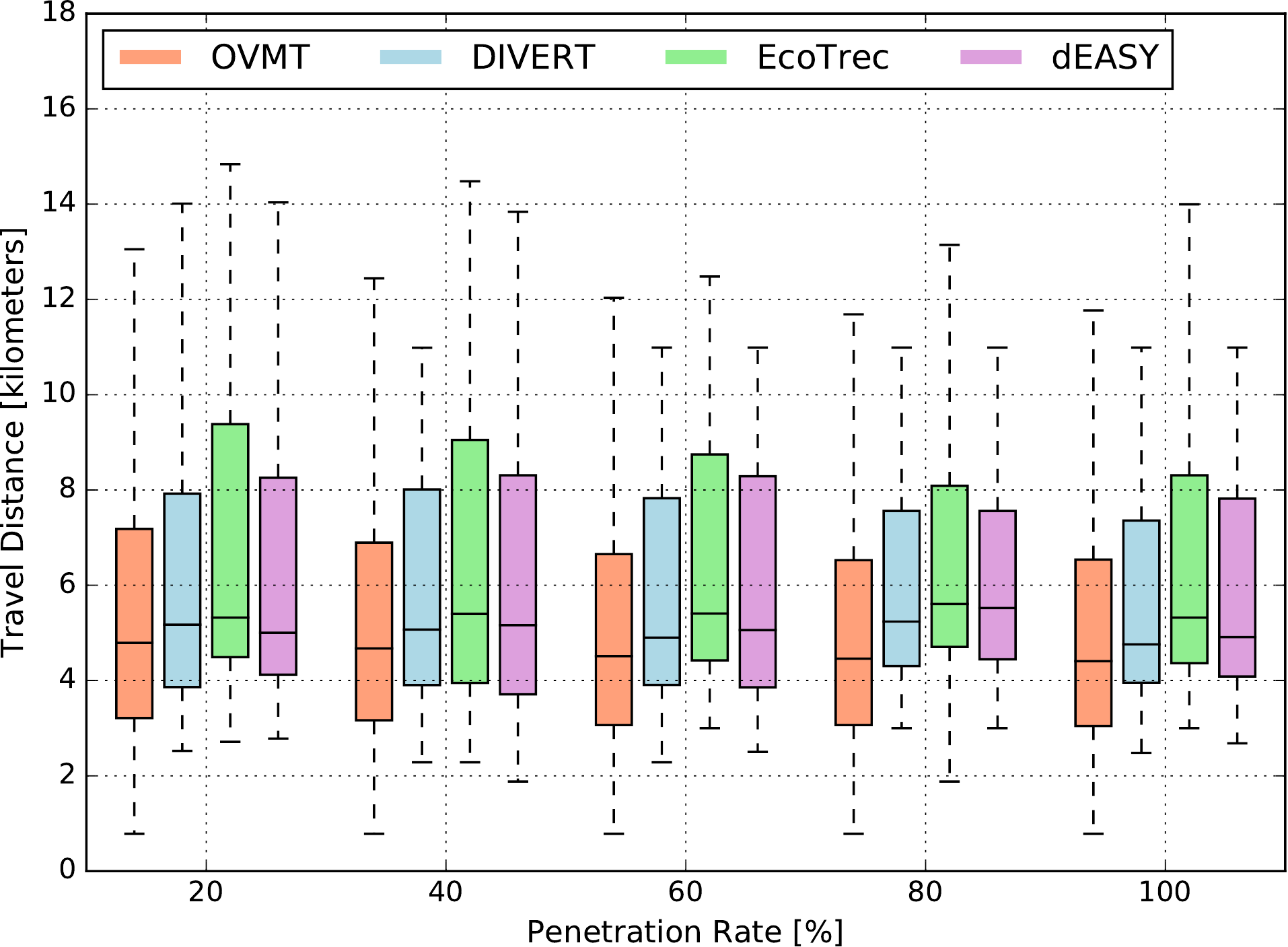}
	}
	\subfigure[Congestion time loss.]{
	\label{fig:congestion_time}
	\includegraphics[width=.47\textwidth]{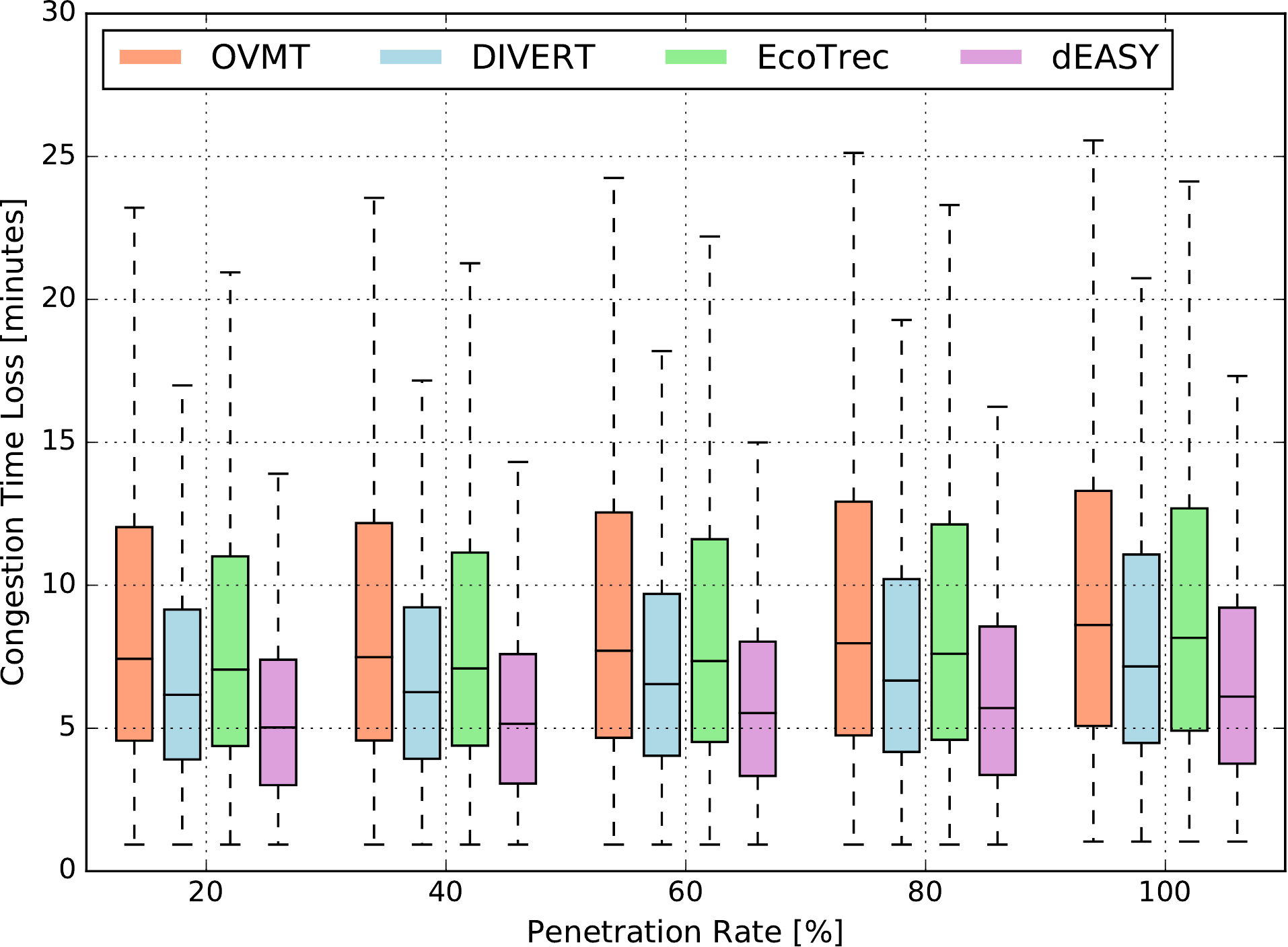}
	}
	\subfigure[CO$_2$ emission.]{
	\label{fig:co2_emission}
	\includegraphics[width=.47\textwidth]{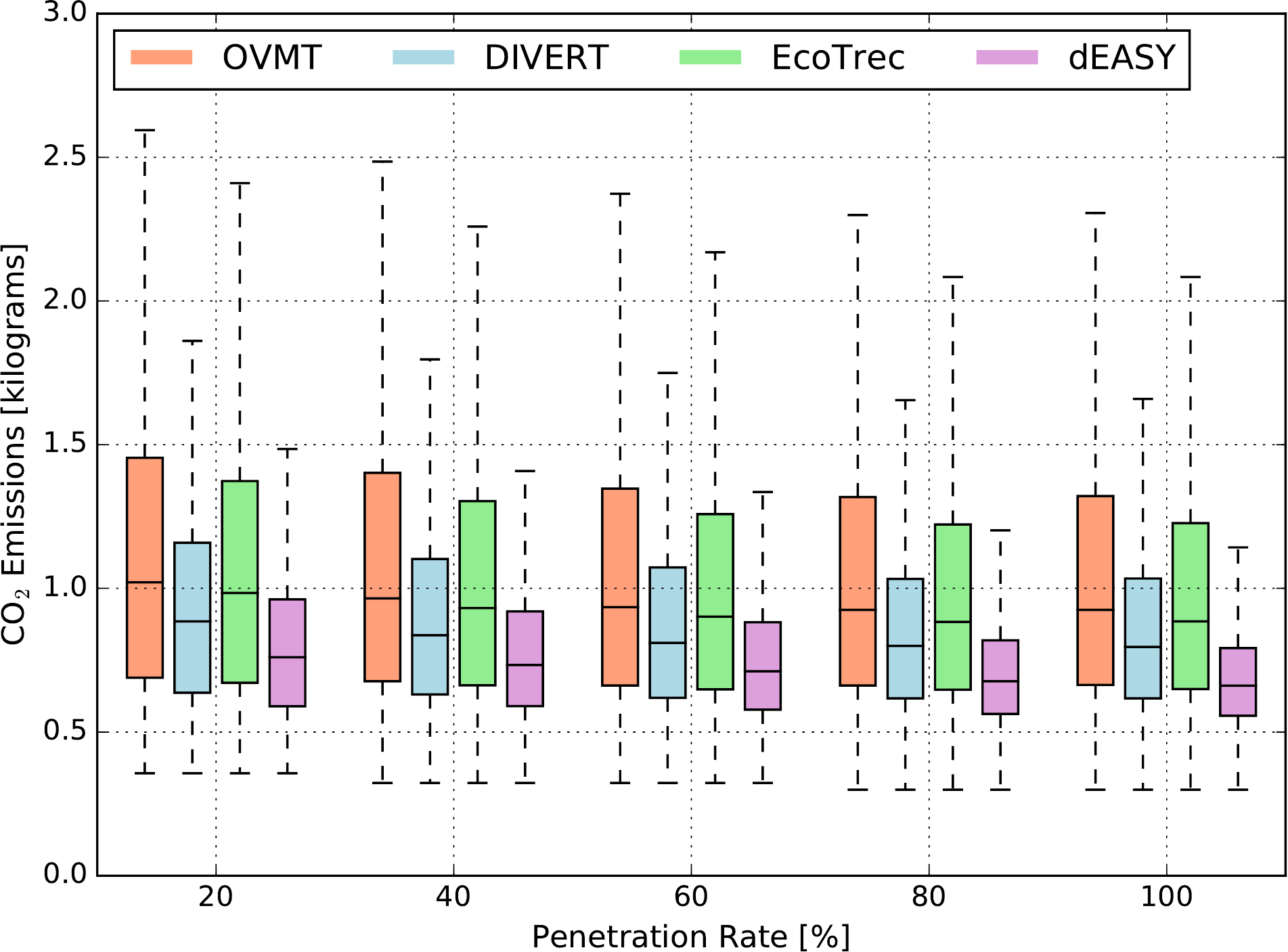}
	}
	\subfigure[Planning time index.]{
	\label{fig:planning_time_index}
	\includegraphics[width=.47\textwidth]{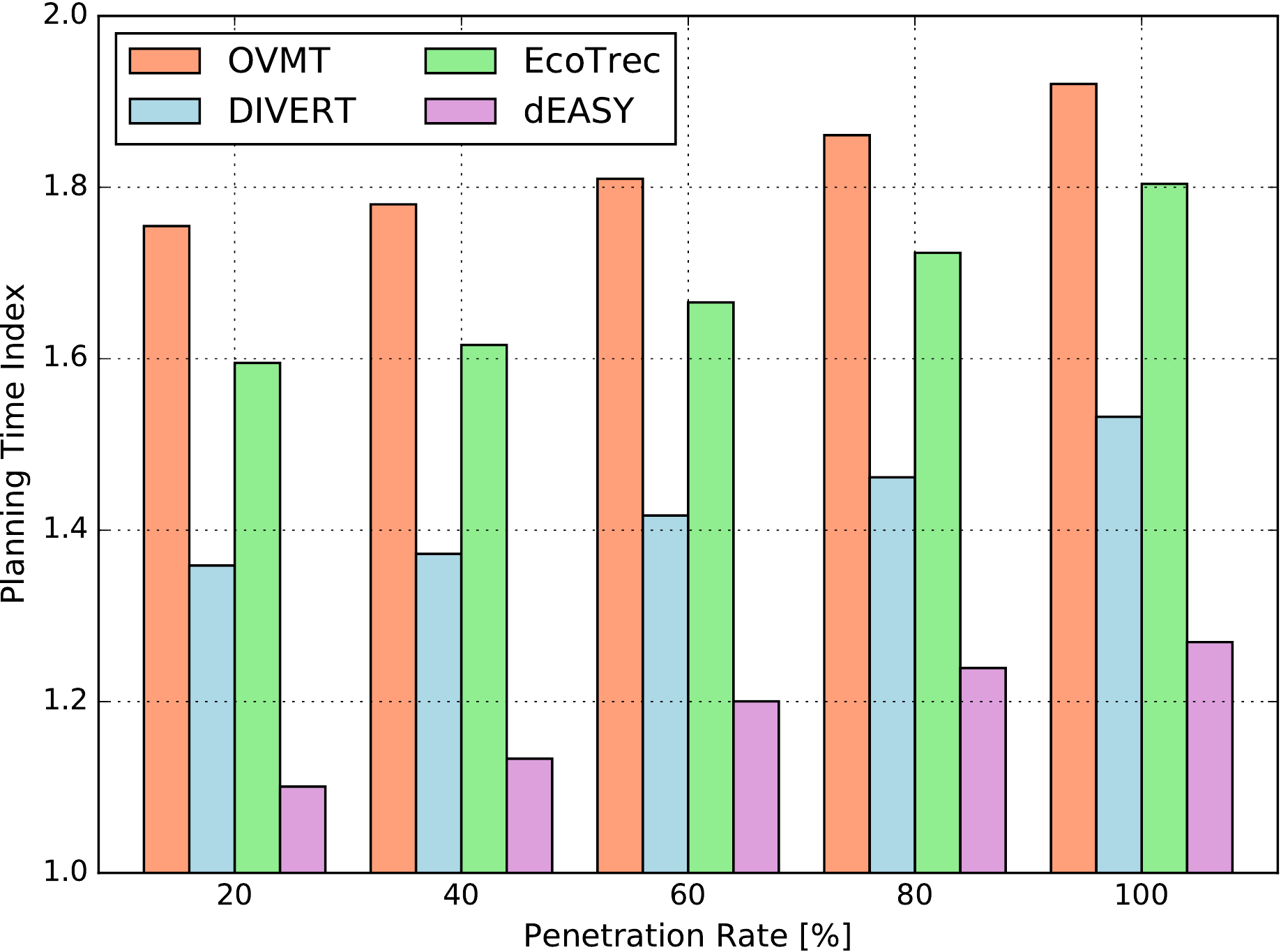}
	}
	\caption{The results of the traffic management assessment.}
	\label{fig:traffic_management}
\end{figure*}

Besides the control channel features and scalability issues, the traffic management also plays an important role in the dEASY system. This subsection compares our proposed system against the DIVERT and EcoTrec systems, as well as OVMT regarding travel time, travel distance, congestion time loss, CO$_2$ emission, and planning time index, as shown in Figure~\ref{fig:traffic_management}. Initially, Figure~\ref{fig:travel_time} shows the result of the travel time for all the penetration rates. The higher the penetration rate, the longer the average travel time for all solutions analyzed. This behavior is in agreement once at high penetration rates the roads become much more congested. As expected, the original trace of K\"oln~(OVMT) has the highest average travel time of around \unit[20]{minutes} in all penetration rate analyzed. Because it keeps the original route throughout the simulation. The EcoTrec system uses a selfish vehicle rerouting that selects an alternative route based on the lower CO$_2$ emissions rate throughout the vehicle's journey. Note that this criterion alone does not present a significant reduction in relation to the travel time metric over the OVMT. In the DIVERT system, vehicles compute an alternative route based on an altruistic routing decision. This means that the alternative route is computed taking into account the routes of surrounding vehicles. The main idea is to avoid a secondary congestion. Thus, it is possible to notice a reduction in the travel time rate of \unit[18]{\%} and \unit[17]{\%} at all analyzed penetration rates, compared with the OVMT and EcoTec, respectively. The dEASY system applies an altruistic routing decision and an entropy-based shortest-path approach to assist in route planning. This combination enables it to outperform all its competitors. By analyzing numerically, dEASY has an average reduction in the rate of around \unit[26]{\%}, \unit[25]{\%}, and \unit[14]{\%} compared with the OVMT, EcoTrec, and DIVERT, respectively.

Figure~\ref{fig:travel_distance} shows the result of the travel distance for all penetration rates. As expected, the systems that apply some vehicle routing algorithm will generally have a greater distance traveled in comparison to the OVMT. Because the OVMT's routes were generated based on the shortest path between the origin and the destination of each vehicle. The EcoTrec system has an average increase in distance traveled of \unit[20]{\%} compared with the OVMT, while the DIVERT system had an average increase of \unit[17]{\%} in distance traveled compared with the OVMT. Furthermore, a decrease in the distance traveled can be observed when comparing with EcoTrec. This is due to the fact of applying the altruistic routing decision in choosing an alternative route. The dEASY system has a higher average distance traveled than the OVMT. However, it is slightly below the average value of the EcoTrec system and practically the same as the average value of the DIVERT system.

Another metric evaluated is the congestion time as the function of the penetration rate Figure~(\ref{fig:congestion_time}). Despite the increase in traveled distance of all analyzed systems, note that there was no increase in time wasted on congestions in relation to OVMT. This was already expected, since all systems apply some mechanism to bypass congestion areas. Such a mechanism is directly associated with the better distribution of vehicles on public roads. The results of the congestion time loss metric in both OVMT and EcoTec are visually the same. This can be attributed to the fact that in the EcoTrec system, the vehicles with similar destination can be moved to the same alternative route, thus, causing a secondary congestion. DIVERT reduced the time lost in congestion by \unit[26]{\%} and \unit[25]{\%} compared with the OVMT and EcoTrec, respectively. Nonetheless, the alternative routes chosen by the dEASY system have the lowest traffic congestion level compared with the other three systems. Considering numbers, dEASY achieves an average reduction in the rate of about \unit[50]{\%}, \unit[48]{\%}, and \unit[30]{\%}, compared with the OVMT, EcoTrec, and DIVERT, respectively. As previously shown, the dEASY system has the lowest network overhead compared with the other two systems. This fact contributes to the information reaching the largest number of participants, and thus, collaborating in the best planning of alternative routes.

Figure~\ref{fig:co2_emission} shows the impact of the penetration rate on the CO$_2$ emission. OVMT has the highest average CO$_2$ emission at all penetration rates. This is due to the fact that the vehicles spend a long time stuck in congestions, as observed in Figure~\ref{fig:congestion_time}. In addition to that, the number of accelerations and decelerations caused by congestions also tends to be greater than in free-flow. These two observations help to explain the reason for the higher CO$_2$ emissions. The results of the congestion time loss and travel time metrics (Figures~\ref{fig:congestion_time} and~\ref{fig:travel_time}) in both OVMT and EcoTec are visually the same. As a consequence, both also have similar behavior in the CO$_2$ emission metric. The DIVERT system has low congestion levels over the OVMT and EcoTrec, due to the vehicles computing an alternative route based on an altruistic routing decision, consequently, it also has the lowest CO$_2$ emission compared with them. dEASY has the lowest CO$_2$ emission average at all analyzed penetration rates. Its average reduction in the rate was around \unit[33]{\%}, \unit[32]{\%}, and \unit[10]{\%} compared with the OVMT, EcoTrec and DIVERT, respectively. Because of that, it holds true that vehicles using dEASY had the least time spent in congestion, thus avoiding the constant braking and acceleration usually needed during traffic congestion.

Figure~\ref{fig:planning_time_index} shows the reliability measure of the planning time index. This index estimates how much additional time should be reserved for a trip relative to free-flow travel time. In other words, it indicates how much of the total time should be added to ensure on-time arrival to the destination on 95\% of the time. As expected, the higher the penetration rate in the simulations, the higher the planning time index for all analyzed systems. This happens because the time spent in congestion also increases, as illustrated in Figure~\ref{fig:congestion_time}. The planning time index is higher in OVMT, followed by EcoTrec, DIVERT, and dEASY. This result is in agreement with the previous ones, such as because of both the travel time and the congestion time. In both cases, they follow the same behavior as shown in Figure~\ref{fig:planning_time_index}. 

The main lesson learned from the results is that the altruistic rerouting approach, together with the entropy-based shortest-path mechanism, helps to improve in the planning of alternative routes.

\section{Conclusion and Future Research} 

This paper presented dEASY, a \textbf{d}istributed v\textbf{E}hicle tr\textbf{A}ffic management \textbf{SY}stem. dEASY is an infrastructure-less and cost-effective system that applies real-life settings for efficient vehicle traffic rerouting. Another advantage of the proposed system is its low network overhead. This system employs an egocentric betweenness measure together with the radio propagation model, as a vehicle selection mechanism, and also an altruistic routing decision and an entropy-based shortest-path approach for vehicle traffic management. A set of simulation experiments, in a real urban scenario, have been performed to analyze the performance of the dEASY system against two other systems: EcoTrec and DIVERT. The experiment results were divided from three perspectives: \textit{(i)} control channel assessment; \textit{(ii)} scalability assessment; and \textit{(iii)} traffic management assessment, and our proposed system outperformed in all the analyzed perspectives.

Our proposed system provides a valuable and timely contribution towards a distributed and infrastructure-less for intelligent and connected transportation systems.

As future work, we intend to apply two essential characteristics of the driver such as mobility pattern, and social relationship for rerouting decisions, to make the most of the available road infrastructure and consequently decrease the time wasted in congestion.

\section*{Acknowledgment}
This work was partially supported by the grants 2015/25588-6, 2016/24454-9, and 2018/02204-6 of the Sao Paulo Research Foundation (FAPESP), and also by the National Council for Scientific and Technological Development (CNPq 401802/2016-7). This work is part of the INCT project named the Future Internet for Smart Cities (CNPq 465446/2014-0, CAPES 88887.136422/2017-00 and FAPESP 2014/50937-1).



\bibliographystyle{elsarticle-num}

\bibliography{bibliography}

\end{document}